\documentclass[preprint,12pt,3p]{elsarticle}

\usepackage{graphicx}
\usepackage{subcaption}

\usepackage{amssymb}
\usepackage{mathtools}

\usepackage{amsmath,amssymb}
\usepackage{xcolor}
\usepackage{mathrsfs}
\usepackage{esint}
\usepackage{siunitx}
\usepackage[percent]{overpic}

\usepackage[numbers]{natbib}
\usepackage[hidelinks]{hyperref}

\newcommand{\diff}{\; \mathrm{d}}

\newcommand*{\dnach}[2]{\frac{\partial {#1}}{\partial {#2}}}

\newcommand{\etal}{et al.\@ }
\newcommand*{\vecu}[1]{\ensuremath{\mathbf{#1}}}

\newcommand*{\dnachc}[3]{\ensuremath{\left(\frac{\partial {#1}}{\partial {#2}}\right)_{#3}}}
\newcommand*{\mat}[1]{\ensuremath{\frac{\mathrm{D} {#1}}{\mathrm{D} t}}}
\newcommand*{\ui}{\ensuremath{\vecu{u}_i}}
\newcommand*{\uf}{\ensuremath{\vecu{u}_f}}
\newcommand*{\ut}{\ensuremath{\tilde{\vecu{u}}_f}}
\renewcommand*{\div}{\nabla \cdot}
\newcommand*{\pf}{\ensuremath{p^{n+1}_f}}
\newcommand*{\pni}{\ensuremath{p^{n+1}_i}}
\newcommand*{\nf}{\ensuremath{\vecu{n}_f}}
\newcommand*{\hatnf}{\ensuremath{\hat{\vecu{n}}_f}}
\newcommand*{\rhoaf}{\rho^a_f}

\makeatletter
\def\ps@pprintTitle{%
 \let\@oddhead\@empty
 \let\@evenhead\@empty
 \def\@oddfoot{}%
 \let\@evenfoot\@oddfoot}
\makeatother
\usepackage{fancyhdr}
\fancypagestyle{pprintTitle}{%
\lhead{}\chead{}\rhead{}
\lfoot{\tiny{UK Ministry of Defence \textcopyright ~Crown Owned Copyright 2026/AWE}}\cfoot{\thepage}\rfoot{}

}

\begin{document}

\begin{frontmatter}

\title{A Semi-Implicit All-Mach Numerical Method for Implicit Large Eddy
Simulations}


\author{Simon D. Wilkinson}
\ead{simon.wilkinson@awe.co.uk}

\address{AWE Nuclear Security Technologies, Aldermaston, Reading, RG7 4PR, UK}

\begin{abstract}
Implicit Large Eddy Simulation (ILES) is a well established approach for
modeling high-Reynolds-number turbulent flows by numerical solution of the
inviscid Euler equations in 3D. These turbulent flows cover a wide range of Mach
numbers, including within individual simulations. For example,
Richtmyer-Meshkov instability simulations transition from a high-Mach
compressible regime to a low-Mach quasi-incompressible regime during the
late-time evolution.
These flows motivate the development of a semi-implicit all-Mach
numerical method,
which is subject to a CFL stability constraint corresponding to the fluid
velocity, but independent of the speed of sound, and can be applied efficiently
and accurately across the whole spectrum of Mach numbers. Furthermore, finite-volume
discretization is chosen to
maintain strict conservation and ensure robust and accurate capturing of strong shocks.
The method is validated in both high- and low-Mach regimes.
Validation is also presented for stratified atmospheres in the presence of
gravity, and the method is thus shown to be applicable to buoyancy driven
flows.
We present favorable comparison against previously published ILES simulations of
multimode Richtmyer-Meshkov instability from the `$\Theta$-group'
collaboration, Thornber et al.,
\emph{Physics of Fluids}, \textbf{29} (10), 2017 \cite{thornber2017late}.

\vspace{0.5cm}

\noindent \small{\textbf{UK Ministry of Defence \textcopyright ~Crown Owned Copyright 2026/AWE}}

\end{abstract}

\end{frontmatter}

\section{Introduction}
\label{sec:introduction}

Compressible, inviscid, high-Mach flow is often modeled using fully
explicit, Eulerian, finite-volume or finite-difference numerical methods
\cite{Toro2009,barton2019interface,wilkinson2022thermal,leveque1992numerical}.
Such methods are well understood for modeling of shock waves in generic
materials
and have also proved successful for modeling turbulence by way of implicit
large eddy simulation (LES)
\cite{thornber2017late,grinstein2007implicit,thornber2016impact}. These methods
are popular on account of their simple finite-volume discretization which
facilitates their implementation and enables coupling with other physics models.

However, the performance of these methods is limited in correspondence with the
speed of sound of the material. The CFL stability criterion \cite{courant1928partiellen}
imposes a restriction on the timestep,
\begin{align}
  \Delta t \lesssim \frac{\Delta x}{|\vecu{u}|+c}, \label{eq:cfl_sound}
\end{align}
where $\Delta x$ is the cell size, $\vecu{u}$ is the flow velocity and $c$ is
the speed of sound.
As such, there is a maximum timestep,
\begin{align}
  \frac{\Delta x}{c},
\end{align}
which cannot be overcome even in the limit of small Mach number.

For high-Mach flow, where the flow velocity is of similar magnitude
to the speed of sound, this
limitation is not overly restrictive, because the timestep limit associated with
the advection of material is not significantly longer than that associated with
the stability criterion.
For low-Mach flow, on the other hand, the CFL stability criterion
restricts the timestep much more than is necessary to maintain accuracy of
bulk advected variables. For this reason, low-Mach flow is often modeled using
so-called low-Mach methods which yield better
computational performance in this regime. The key characteristic is that the
stability criterion of the method is not dependent on the speed of sound, and
is instead of the form
\begin{align}
  \Delta t \lesssim \frac{\Delta x}{|\vecu{u}|}. \label{eq:cfl_advection}
\end{align}

The implication is that the number of computational timesteps required to
evolve the state for a low-Mach method relative to a method subject
to \eqref{eq:cfl_sound} is
\begin{align}
  \frac{|\vecu{u}|}{|\vecu{u}| + c} = \frac{1}{1 + 1/M},
\end{align}
for a flow with Mach number $M$. For example, a flow at Mach number $0.1$ can
be simulated with roughly an order of magnitude fewer timesteps than would be
required with a method subject to \eqref{eq:cfl_sound}.

A survey of low-Mach methods is provided by Almgren \etal \cite{almgren2006low}.
The simplest example is the incompressible Navier-Stokes equations, where the
velocity divergence is set to zero.
There are various methods that allow for density variations within specific
conditions, for example the Boussinesq approximation can be used to
capture the effects of buoyancy within a constant density domain
\cite{boussinesq1903theorie}, however this is subject to the requirement that
the relative density variation is small, $\Delta \rho \ll \rho$.
Another example is methods reliant on the anelastic continuity equation which
is suitable for flows where density variations occur within a stratified
atmosphere.

All-Mach methods are distinct from low-Mach methods in that they
combine the ability to model strong shock waves of CFL-limited explicit
methods with the stability criterion of low-Mach methods
\eqref{eq:cfl_advection}.

There are many scenarios where such a method offers distinct advantages.
Firstly, simulations for which the Mach number varies spatially within the domain. This
commonly occurs in multi-material simulations, for example an air-water
flow where a shock wave is present in the air and simultaneously the flow in
the water is well characterized as being incompressible
\cite{tryggvason2011direct}. For
this publication, however we restrict ourselves to single material simulations
using the ideal gas equation of state.

Secondly, flows that cover a range of Mach numbers in time.
A typical characteristic of turbulent flow is that the kinetic energy gradually
decays with time as energy is dissipated in the form of heat \cite{thornber2016impact}.
Correspondingly, the Mach number decreases over the course of the simulation,
and the regime of the flow transitions from a high-Mach regime, where
compressibility of the material is considerable, to a low-Mach regime
where the compressibility of the material can be neglected.

One example of this is Richtmyer-Meshkov instability. The turbulence is generated as
a shock wave transits a material interface, and so any suitable numerical method must
capture shock waves accurately. On the other hand, evaluation of the late-time mixing
layer width requires simulation at very low Mach numbers.
Oggian \etal \cite{oggian2015computing} present a method involving switching
from a compressible to incompressible method during a simulation.
However, this approach is not ideal. If the compressible method is used too late, the
associated computational cost saving will be reduced, and if the incompressible
method is used too soon, then this can be inaccurate owing to the
divergence of the flow field prior to switching.
A single method capable of operating efficiently and accurately
across the entire Mach number spectrum is preferable.
We will present results for late-time Richtmyer-Meshkov instability
in this work.

Thirdly, a scenario where all-Mach methods are a natural choice is the
modeling of flows in a stratified atmosphere. Significant movement of material
in the vertical direction within a stratified domain will naturally lead to
compression / expansion of material, and this occurs even in the context of
low-Mach number flow. In some scenarios such flows can be modeled well
using the Boussinesq approximation, however this does not allow expansion and
compression of the material as it moves, and does not permit the use of
generic, complex equations of state.
The method presented here can be applied to stratified, low-Mach flow
directly, providing good computational performance with no additional
approximation necessary, and can trivially be applied to any material model. An
example of a numerical method developed for this type of flow is that presented by
Restelli and Giraldo \cite{restelli2009}. We validate the method for this type of flow using the
example of a buoyant bubble in a stratified atmosphere
\cite{robert1993}.

Kwatra \etal \cite{Kwatra2009} presented an all-Mach method for evolving
a conservative finite-volume discretization of the Euler equations.
The basic principle is to evolve the state by
way of an explicit advection scheme coupled with an implicit solution for the
pressure field. As such it is possible to construct a fully conservative
method, hence ensuring that the correct shock propagation
speed is captured \cite{leveque1992numerical}. The advection of material is modeled through an explicit (local)
update, and the associated timestep is thus constrained by the advection
velocity for stability. Compression waves, on the other hand, are captured
within the implicit solution of the pressure field. This part of the method is
therefore unconditionally stable, and does not constrain the timestep beyond
the constraints imposed by the advection.
Kanarska \etal have applied this method for modeling atmospheric flows
\cite{Kanarska2020}. 

The method presented here follows the work of Jemison \etal
\cite{jemison2014compressible} and Kuhn and Desjardins
\cite{kuhn2021all} in splitting
the advection step into a predictor step and a subsequent corrector step.
This has been found to significantly improve the quality of the solution in the
very low-Mach regime.
Further details of the method have been refined and simplified with the
specific aims of accurate, practical implicit LES modeling of turbulence in
stratified flows and in the presence of shock waves. The model has been
validated for the modeling of shocks, isotropic turbulence, buoyancy driven
turbulence, and Richtmyer-Meshkov instability.

Fuster and Popinet \cite{fuster2018all} have applied all-Mach methods to multi-material simulations
and have developed a model including surface tension effects.
Urbano \etal \cite{urbano2022semi} and Demou \etal \cite{demou2022pressure}
are examples of similar methods that differ from the method presented here in
that they are constructed around the primitive variables and thus do not impose
strict conservation in the presence of strong shocks.

Dumbser \etal \cite{dumbser2015,dumbser2016} have presented all-Mach
methods subject to a CFL condition associated with the fluid velocity. These
methods are implemented on a staggered grid, with the velocities defined on the
cell faces, while density and energy is defined at the cell centers. The scheme
presented here is fully cell-centered and thus has the same conservation
properties of other finite-volume schemes.
Boscheri \etal \cite{boscheri2020,boscheri2021} build on the work of Dumbser
and extend the method for the Navier-Stokes equations. Busto and Dumbser \cite{busto2022}
apply a similar method to the shallow water equations.

The work presented here is the first example we are aware of where an all-Mach
method is applied to implicit LES modeling of turbulent flows whose Mach
number evolves substantially during the simulation.
In particular,
late-time Richtmyer-Meshkov calculations which necessarily have to bridge
the gap from the compressible shock wave to low-Mach turbulent
mixing \cite{oggian2015computing}.
Such flows require accurate
shock capturing at early times while remaining computationally efficient in the
low-Mach regime at late times.
To the author's knowledge, this is the first
demonstration of a conservative all-Mach method specifically developed and
validated for ILES applications.

The principal contributions of this work are:
\begin{enumerate}
  \item Development of a conservative cell-centered all-Mach finite-volume
  method for the Euler equations.
  \item A predictor-corrector formulation which recovers accurate low-Mach
  behavior while retaining shock-capturing capability.
  \item A pressure interpolation strategy that yields a well-balanced
  discretization for stratified atmospheres.
  \item Demonstration that the method is suitable for ILES of turbulent flows
  spanning a wide range of Mach numbers.
  \item Application to late-time Richtmyer-Meshkov instability where the
  increasing timestep provides substantial computational savings.
\end{enumerate}

\section{Method}
\label{sec:method}

In this paper we focus on implicit LES (ILES) modeling by way of solution of the
inviscid Euler equations. In conservative
form, and using $\rho$ for density, $\vecu{u}$ for velocity, $E$ for energy density, $p$
for pressure, $\vecu{g}$ for the acceleration due to gravity and $\otimes$ for the outer product, 
the Euler equations are
\begin{align}
  \dnach{}{t} \begin{pmatrix} \rho \\ \rho \vecu{u} \\ E \end{pmatrix}
  + \div \begin{pmatrix} \rho \vecu{u} \\ \rho \vecu{u} \otimes \vecu{u} + p \vecu{I} \\ (E + p) \vecu{u}   \end{pmatrix}
  = \begin{pmatrix} 0 \\ \rho \vecu{g}  \\ \rho \vecu{g} \cdot \vecu{u} \end{pmatrix} \label{eq:euler}
\end{align}
The equations are closed with an equation of state for the pressure in terms of
the specific internal energy, $e$,
\begin{align}
 E &= \rho (e + \frac{1}{2} \vecu{u} \cdot \vecu{u}) \\
 p &= p(\rho, e). \label{eq:eos_generic}
\end{align}
The results presented in this work use the ideal gas equation of state,
\begin{align}
  p = (\gamma - 1) \rho e, \label{eq:ideal_gas}
\end{align}
with adiabatic gamma, $\gamma$, set to be $1.4$ unless stated otherwise. The isentropic
bulk modulus, $K$, is defined to be
\begin{align}
  K = \rho \dnachc{p}{\rho}{S} = \gamma p, \label{eq:isentropic_bulk}
\end{align}
where $S$ is used for entropy.
Note that the method presented here makes no inherent assumptions about the
equation of state.

The algorithm presented here has been implemented for 3D Cartesian structured
meshes within the AMReX framework \cite{amrex}.

We also include in the method generic advected variables $z_n$ which enable
future extensions of the model for more complex physics and more complex
material models. In this work they are used as tracer variables
only. They are advected according to
\begin{align}
  \dnach{z_n}{t} + \vecu{u} \cdot \nabla z_n = 0. \label{eq:advection}
\end{align}
In the following development of the method we recast the advection equation for $z_n$
into conservative form following Johnsen and Colonius \cite{Johnsen2006},
\begin{align}
  \dnach{z_n}{t} + \div (z_n \vecu{u}) = z_n \div \vecu{u}. \label{eq:advection_conservative_form}
\end{align}

Equations \eqref{eq:euler} and \eqref{eq:advection_conservative_form} are discretized
with the finite-volume method. This ensures conservation of mass, momentum and energy which is desirable
for correct capturing of shock waves \cite{leveque1992numerical}. The state is
encoded using cell-centered variables, where for brevity we use a single index,
$i$, to label the cell.
\begin{align}
  \vecu{U}_{i} = \begin{pmatrix} \rho_i \\ \rho_i \vecu{u}_i \\ E_i \\ z_{n,i} \end{pmatrix} \label{eq:conservative_vars}
\end{align}
Variables defined on cell faces are indexed using an index $f$.

The algorithm is laid out below.
\begin{itemize}
\item (Section \ref{sec:time_step}) The first step is to determine a suitable timestep, $\Delta t$, using the flow
velocity field, $\ui$.

\item (Section \ref{sec:predicted_face_velocity}) A velocity field, $\ut$,
the `predicted face velocities' are constructed on the cell faces using
a MUSCL reconstruction of the cell-centered data. Each component of the velocity is defined on the faces
normal to that component only. The procedure for the initialization of this field is
described in section \ref{sec:predicted_face_velocity}. More details of the
MUSCL reconstruction procedure are given in Section \ref{sec:muscl}.

\item (Section \ref{sec:predictor}) The predicted face velocities are used to integrate the density equation and the
pressure advection equation by one timestep.
\begin{align}
  \dnach{\rho}{t} + \div \rho \vecu{u} &= 0 \\
  \dnach{p}{t} + \vecu{u} \cdot \nabla {p} &= 0
\end{align}
The initial density field is defined by $\rho_i$, while the pressure field is
initialized using the conserved variables in $\vecu{U}_i$ and the equation of
state.
The results of the advection are labeled $\rho^a_i$ and $p^a_i$. The advected
density is then interpolated to the cell faces, $\rhoaf$, using a simple
average of the neighboring cells.

\item (Section \ref{sec:linear_system}) The results of the advection step are used to setup a linear
system for the cell-centered pressure field at the end of the timestep, $\pni$,
\begin{align}
  \left(\frac{1}{K^a} - \Delta t^2 \div \frac{1}{\rho^a} \nabla \right) p^{n+1} &=
  \frac{p^a}{K^a} - \Delta t \div \ut.
\end{align}
where $K^a$ is the isentropic bulk modulus evaluated using the advected
variables, $\rho^a_i$, $p^a_i$.
The cell-centered pressure field, $\pni$, is used to construct the
corrected face velocities, $\uf$.
\begin{align}
  \uf = \ut + \Delta t \left( - \frac{1}{\rhoaf} \nabla \pni + \vecu{g} \right)
\end{align}
A face centered pressure field, $\pf$, is constructed using a specific volume
weighted average of $\pni$.

\item (Section \ref{sec:integration})
Fluxes, $\vecu{F}_f$, and source terms, $\vecu{S}_i$, are calculated for each cell face, using a MUSCL reconstruction of
the original state on the faces, $\vecu{\bar{U}}$,
the corrected face velocities, $\uf$, and the solution for the
pressure field, $p^{n+1}$,
\begin{align}
  \vecu{F}_f = \uf \vecu{\bar{U}_f} + \pf \begin{pmatrix} 0 \\ \hatnf \\ \uf \cdot \hatnf \\ 0 \end{pmatrix}, \qquad
  \vecu{S}_i = \begin{pmatrix} 0 \\ \rho_i \vecu{g} \\ \rho_i \vecu{g} \cdot \uf \\ z_{n,i} \div \uf \end{pmatrix},
\end{align}
and source terms where $\nf$ is the unit vector normal to the faces.

Finally, the state is integrated in time using a Runge-Kutta integration, where
the MUSCL reconstruction is repeated for each stage, but the 
$\uf$ and $\pni$ fields are re-used for all stages.

\item (Section \ref{sec:boundary_conditions})
The evaluation of the fluxes on faces at the boundaries of the mesh is
detailed here.

\end{itemize}

\subsection{Stability constraint on the Timestep}
\label{sec:time_step}

The stability of the numerical method is constrained by the stability of the
explicit integration of the advection terms. This corresponds to a stability
constraint on the timestep corresponding to the flow velocity,
\begin{equation}
  \Delta t \le \Delta t_{CFL} \coloneqq \frac{1}{\sum_i \frac{u_i}{\Delta x_i}}
\end{equation}
where the sum is over the spatial dimensions, and $\Delta x_i$ are the
corresponding cell dimensions.

The timestep is furthermore required to be such that the change in velocity
due to gravity over a single timestep is also within the constraints for
stability,
\begin{align}
  \Delta t &\le \frac{\Delta x}{g \Delta t} \;\; \Rightarrow \;\;
  \Delta t \le \Delta t_{Fr} \coloneqq \sqrt{\frac{\Delta x}{g}}
\end{align}
where we have implicitly assumed here that gravity only acts in one dimension.
In terms of the Froude number, $Fr \coloneqq u / \sqrt{gL}$, using domain
height, $L$, and setting $N \coloneqq L / \Delta x$ to be the
number of cells across the height of the domain then this additional
restriction on the timestep is relevant only if
\begin{equation}
  Fr < \frac{1}{\sqrt{N}} \label{eq:froude_restriction}.
\end{equation}
For three dimensional simulations in this low Froude number regime,
the total computational cost to advance the state a unit of time
scales like $\mathcal{O}(N^{3.5})$.
This scaling will continue while \eqref{eq:froude_restriction} is satisfied,
and then, for larger $N$, reverts to the usual $\mathcal{O}(N^4)$ scaling.
It is worth noting that for strong stratification (i.e.\ large changes in
density between neighboring cells) and small Froude number, an explicit
numerical method subject to
\eqref{eq:cfl_sound} may be faster. Specifically, this occurs when
\begin{equation}
  g \gtrsim \frac{c^2}{\Delta x}.
\end{equation}
For Earth's atmosphere, this corresponds to $\Delta x \approx 12\mathrm{km}$.

In practice, we obtain good results by preventing excessively aggressive
lengthening of the timestep and further limit the timestep using the previous
timestep $\Delta t_{prev}$ as follows,
\begin{align}
  \Delta t \gets \max\left(1.05 \Delta t_{prev}, \Delta t_{CFL}, \Delta t_{Fr}, \right).
  \label{eq:time_step_evolution}
\end{align}

\subsection{Predicted Face Velocity Construction}
\label{sec:predicted_face_velocity}

These face velocities are only used for the predictor step and will then be
`corrected' using the pressure field that is obtained by solving the linear
system. As such there is some flexibility in the evaluation of these face
velocities and there are many approaches which will work. Nevertheless the
choice of approach will affect the quality of the solution.

For the high-Mach regime we take inspiration from the physical understanding from the
solution to the Riemann problem
\cite{Toro2009} and use upwinding in flows which suggest the presence of shock
waves (converging flow), and an average in the presence of rarefaction waves
(diverging flow). Furthermore, taking inspiration from the low-Mach
correction presented by Thornber \etal \cite{thornber2008lowmach}, if the flow is in a
low-Mach regime we take the mean of the cell-centered data. This helps to reduce
the dissipation when the Mach number is low. In addition, we demand that in the
context of symmetric flow on either side of the face, the face velocity will be
evaluated to be zero. Due to numerical considerations, this requirement has to extend to situations where
numerical error has introduced very small asymmetries to the data. This ensures
that simulations will maintain symmetry where that is physically expected. An
example of this is in the context of the circular Sod test that is presented
in section \ref{sec:circular_sod}. In addition, we find that the interpolation
to cell faces must be done monotonically -- that is with a resulting value
strictly within the range of velocities either side of the face. This is
necessary to ensure that spurious movement is not generated in the context of
stratified domains. More details of this are discussed in
\ref{sec:pressure_interpolation}.

Finally we look for an expression with these properties which is also
continuous and smooth, avoiding sharp steps that can cause two simulations
with very close starting conditions to diverge in an unphysical manner.

The expression used for the predictor face velocity, $\ut$, is
\begin{align}
  \tilde{u}_{i+1/2} &= \left\{ \begin{array}{ll} s + \min(\alpha M, 1) \mathrm{sign}(s) \frac{u_i - u_{i+1}}{2} & \mathrm{if} \quad u_i > u_{i+1} \\ 
  s & \mathrm{otherwise} \end{array} \right. \label{eq:uface_pred} \\
  \mathrm{where} \qquad s &= \frac{u_i + u_{i+1}}{2}, \qquad M = \frac{|s|}{\min(c_i, c_{i+1})}, \qquad \alpha = 2
\end{align}
where $u_i$, $u_{i+1}$, $c_i$ and $c_{i+1}$ are calculated using the
reconstructed variables, and $c$ is the speed of sound,
\begin{equation}
  c_i^2 = \dnachc{p}{\rho}{S}.
\end{equation}
The variable $M$ can be understood intuitively as corresponding to the Mach
number. The parameter $\alpha$ can be varied, and the optimal value may depend
on the application. Values in the range $[1, 4]$ yield reasonable results. In
this work we settle on the value of 2, which was used for all the simulations
presented in this paper.

Equation \eqref{eq:uface_pred} is not the only possible expression to meet the
requirements detailed above, but it is a simple expression which has proven to
give satisfactory results in practice.

\begin{figure}
  \centering
  \begin{subfigure}[b]{0.8\textwidth}
    \includegraphics[width=\linewidth]{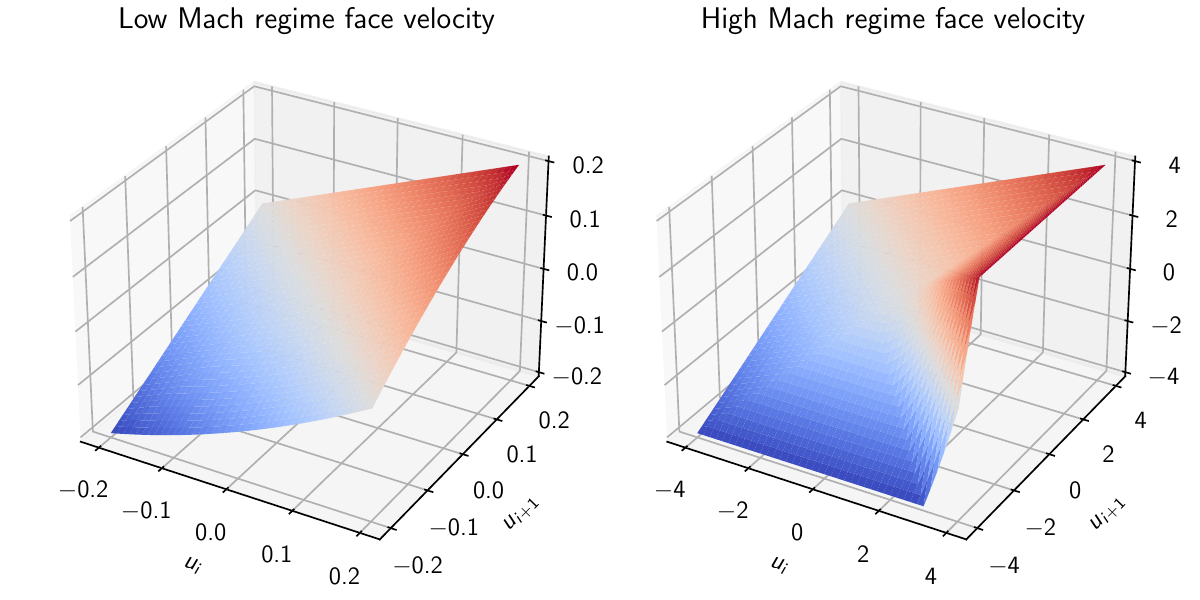}
    \phantomsubcaption
  \end{subfigure}
  \begin{subfigure}[b]{0.8\textwidth}
    \includegraphics[width=\linewidth]{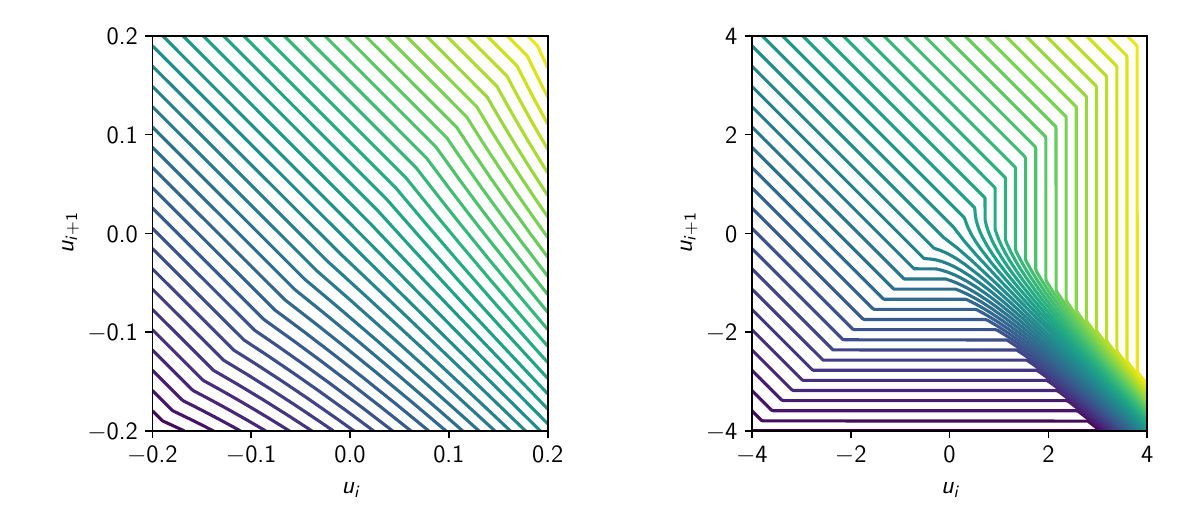}
    \phantomsubcaption
  \end{subfigure}
  \caption{An illustration of equation \eqref{eq:uface_pred} showing the
  predicted face velocity, $\tilde{u}_{i+1/2}$, as a function of velocity from
  the neighboring cells, $u_i$ and $u_{i+1}$. The plots are generated assuming
  $c_i = c_{i+1} = 1$, so the units on the plots can be interpreted as Mach
  numbers.}
  \label{fig:face_velocity}
\end{figure}

Figure \ref{fig:face_velocity} shows the face velocity, $\tilde{u}_f$ as
a function of the velocity on either side of the interface. The figure shows that the
expression \eqref{eq:uface_pred} has the desired properties: in the low-Mach
regime it simply takes the average while smoothly blending into an upwind
approach for converging flow in the high-Mach regime.

\subsection{Explicit Predictor Advection Step}
\label{sec:predictor}

The predictor advection step is required to construct the linear system which
is discussed in section \ref{sec:linear_system}.

We are integrating the equations
\begin{align}
  \dnach{\rho}{t} + \div \rho \vecu{u} &= 0 \\
  \dnach{p}{t} + \vecu{u} \cdot \nabla {p} &= 0, \label{eq:p_advection}
\end{align}
for the advection of the density and pressure fields using
the predictor face velocity field, $\tilde{u}_f$, to calculate the
fluxes.
The pressure, $p_i$, is initially calculated from the conserved variables
\eqref{eq:conservative_vars} using the equation of state.
Equation \eqref{eq:p_advection} is cast into conservative form just as done
previously for \eqref{eq:advection_conservative_form}.
The derivatives are discretized into finite-volume form,
\begin{align}
\vecu{U}_i(t) &= \begin{pmatrix} \rho \\ p \end{pmatrix}, \nonumber\\
  \frac{\partial}{\partial t} \vecu{U}_i 
    &= \frac{1}{V_i} \sum_{f\in \partial V_i} \nf \cdot \ut
    \vecu{\bar{U}}_f(\vecu{U}_i, \vecu{U}_{i+1})
      + \begin{pmatrix} 0 \\ p \div \ut \end{pmatrix},
      \label{eq:finite_volume_advection}
\end{align}
where the sum is over all faces, $f$, adjacent to the cell, $i$, $\nf$ is an
outward vector normal to the face, $f$, with magnitude corresponding to the
area of the face, and $V_i$ is the volume of the cell.
The divergence of the predictor face
velocity field is calculated using the same finite-volume discretization.
$\vecu{\bar{U}}_f$ is used to denote a MUSCL reconstruction of the state on the face.
The integral with respect to time is done using a two stage Runge-Kutta method
(explicit midpoint method). Note that we only construct a single predictor face
velocity field, $\ut$, which is reused for the two stages of the Runge-Kutta
integration. The MUSCL reconstruction is redone for each stage of the
integration and is done using the Kim-Kim 5\textsuperscript{th} order limiter \cite{KimKim2005}. The
Runge-Kutta method is detailed explicitly in section \ref{sec:integration}.

Equations \eqref{eq:finite_volume_advection} are integrated a single timestep,
$\Delta t$, in time, and the results are labeled $\rho^a$ and $p^a$.

Finally the advected isentropic bulk modulus, $K^a$, is evaluated using the equation state
and the advected density and pressure,
\begin{align}
  K^a = K(\rho^a, p^a).
\end{align}

We note that for the ideal gas equation of state, the evaluation of the bulk
modulus reduces to simply, $K^a = \gamma p^a$, 
and the result for $K^a$ is thus identical to what would be obtained if the
bulk modulus was instead evaluated using the initial data and then advected
using the given scheme described above.
More complex equations of state may break this equivalence.

\subsection{Implicit Calculation of Velocity and Pressure fields}
\label{sec:linear_system}

In this section, the evaluation of the corrected velocity and pressure fields, $\uf$ and
$\pf$, is presented. These fields are evaluated implicitly, and thus form a
mutually consistent state. The divergence of the velocity field corresponds to
the pressure change, and likewise the pressure gradient corresponds to the
acceleration. These fields will be used to evaluate fluxes and integrate the
equations as described in section \ref{sec:integration}.

The stability properties of the method hinge on the fact that the pressure
field used in evaluation of the fluxes is an implicit evaluation of the
pressure field at time $t_{n+1}$. This part of the method is based on the
method presented by Kwatra \cite{Kwatra2009}, and uses ideas from
\cite{jemison2014compressible,kuhn2021all}.

The evolution equation for the pressure is
\begin{align}
  \mat{p} &= - K \div \vecu{u}, \label{eq:p_evo}
\end{align}
where $K$ is the isentropic bulk modulus, \eqref{eq:isentropic_bulk}.
Note that equation \eqref{eq:p_evo} corresponds to the first law of
thermodynamics; it is equivalent to equations \eqref{eq:euler} with the
equation of state.

The pressure field at the end of the timestep, $p^{n+1}$, is evaluated using the result of
the previous integration of the advection equation and then integrating the
term on the right hand side of \eqref{eq:p_evo} with respect to time,
\begin{align}
  p^{n+1} &= p^{a} - \Delta t K^a \div \uf,
\end{align}
using a velocity field, $\uf$, representative of the velocity field at the end
of the timestep that is constructed from the predicted velocity field, $\ut$, from
section \ref{sec:predicted_face_velocity}, and the advected densities,
$\rho^a$,
\begin{align}
  \uf &= \ut + \Delta t \left(-\frac{1}{\rho^a} \nabla p^{n+1} + \vecu{g}\right)
  \label{eq:u_correction}
\end{align}
and is itself dependent on the pressure field, $p^{n+1}$. We thus have an
implicit system for the pressure,
\begin{align}
  \frac{p^{n+1} - p^a}{\Delta t} &= -K^a \div \left[ \ut - \Delta t \frac{1}{\rho^a} \nabla p^{n+1} \right],
    \label{eq:helmholtz_div}
\end{align}
which after discretization constitutes a linear system.
The gravity term has been dropped on the assumption that the gravity force is
divergence free. However we note here that in the vicinity of reflective boundaries
this term must be dealt with carefully. This is discussed further in section
\ref{sec:boundary_conditions}.

The density field $\rho^a$ needs to be evaluated on cell faces, and is obtained
from a straightforward averaging process,
\begin{align}
  \rho^a_f = \frac{\rho_i + \rho_{i+1}}{2}, \label{eq:rho_face}
\end{align}
where face, $f$, here refers to the cell face with neighboring cells, $i$ and
$i+1$.
Equation \eqref{eq:helmholtz} is an
inhomogeneous Helmholtz equation, and there are standard methods for its
solution available in many libraries such as AMReX \cite{amrex} and HYPRE
\cite{hypre}. We will therefore not go into detail on the method of its
solution. 
Rearranged to be in the standard form used by AMReX, it reads
\begin{align}
  \left(\frac{1}{K^a} - \Delta t^2 \div \frac{1}{\rho^a} \nabla \right) p^{n+1} &=
  \frac{p^a}{K^a} - \Delta t \div \ut. \label{eq:helmholtz}
\end{align}

After the solution of the linear system for the cell-centered pressure field,
$p^{n+1}$, we must also construct the corrected face velocity field, $\uf$
using \eqref{eq:u_correction} which will be required for the evaluation of
fluxes in the next section. This correction procedure differs from the original
method of Kwatra \etal \cite{Kwatra2009} but is essential for recovering smooth solutions
in the low-Mach number limit. In particular, in the limit
of low compressibility, (in other words large $K$), then the corrected
velocities will tend towards a divergence free field, and the pressure will
start to behave, not as a prediction for the actual pressure at the next timestep,
but rather as a Lagrange multiplier that constrains the amount of
divergence which can be present in the resulting velocity field, $\uf$.

The final step is to interpolate the cell-centered pressure field, $p^{n+1}$, to
the faces for the subsequent evaluation of fluxes. This is done using a
specific volume weighted average for each face, $f$, with neighboring cells
$i$ and $i+1$,
\begin{align}
  p_f &= \frac{\rho_{i+1} p^{n+1}_{i} + \rho_i p^{n+1}_{i+1}}{\rho_i + \rho_{i+1}}.
\end{align}

Kwatra \etal \cite{Kwatra2009} give
an argument as to why this is the best way to interpolate the pressure. 
Further to this, we find that this approach to interpolation of the pressure is essential when
modeling stratified fluids in the presence of gravity, and ensures the method
is `well-balanced' \cite{MichelDansac2025,Thomann2020}. This means that there
exists a stable state where the forces of gravity and the pressure
gradient cancel exactly. More details of this are given in
\ref{sec:pressure_interpolation}.

\subsection{Runge-Kutta integration}
\label{sec:integration}
Using the face centered, corrected velocity field, $\uf$ and the face centered pressure field
$\pf$ originating from the solution of the linear system described in
section \ref{sec:linear_system} we are now ready to integrate the Euler
equations \eqref{eq:euler} and supplementary advection equations
\eqref{eq:advection_conservative_form}.
The equations are discretized into standard finite-volume form
\begin{align}
\vecu{U}_i(t) &= \begin{pmatrix} \rho \\ \rho \vecu{u} \\ E \\ z_n \end{pmatrix}, \nonumber\\
  \frac{\partial}{\partial t} \vecu{U}_i 
    &= \frac{1}{V_i} \sum_{f\in \partial V_i} \left[ \nf \cdot \uf \vecu{\bar{U}}_f(\vecu{U}_i, \vecu{U}_{i+1})
      + \pf \begin{pmatrix} 0 \\ \nf \\ \uf \cdot \nf \\ 0 \end{pmatrix} \right]
      + \begin{pmatrix} 0 \\ \rho_i \vecu{g} \\ \rho_i \vecu{g} \cdot \uf \\ z_{n,i} \div \uf \end{pmatrix},\label{eq:finite_volume_euler}
\end{align}
where we again use $\vecu{\bar{U}}_f$ to denote a MUSCL reconstruction of the state on the face.
More details of this are in section \ref{sec:muscl}.
The ordinary differential equation, \eqref{eq:finite_volume_euler}, is
discretized in
time and integrated numerically via the two stage explicit midpoint
Runge-Kutta method \cite{butcher2008}. Note that the algorithm we use here is not
strictly a two stage Runge Kutta method, since we only determine the face
velocities and face pressure once per timestep.
\begin{align}
  \uf, \pf &= \mathcal{A}(\vecu{U}^n) \\
  \vecu{U}^{(1)} &= \vecu{U}^n + \Delta t \mathcal{L}(\vecu{U}^{n}, \uf, \pf) \\
  \vecu{U}^{n+1} &= \frac{1}{2} \vecu{U}^n + \frac{1}{2} \vecu{U}^{(1)} +
  \frac{\Delta t}{2} \mathcal{L}(\vecu{U}^{(1)}, \uf, \pf)
\end{align}
where we use $\mathcal{L}(\vecu{U})$ to represent the right hand side of equation
\eqref{eq:finite_volume_euler}, and $\mathcal{A}(\vecu{U})$ to represent the
procedure defined in section \ref{sec:predictor} and \ref{sec:linear_system},
and $\vecu{U}^{n}$ to represent the $n^{th}$ timestep of the solution.

\subsection{MUSCL reconstruction}
\label{sec:muscl}
Following other implementations of the MUSCL method we perform the
reconstruction on the primitive variables rather than the conservative
variables. This is preferable since it ensures the correct linear
reconstruction in the presence of contact waves across which the primitive
variables, velocity, $\vecu{u}$, and pressure, $p$ are constant, but the
momentum and energy will differ. See for example Johnsen and Colonius
\cite{Johnsen2006}.

The primitive variables are calculated using the equation of state
\eqref{eq:eos_generic} and are defined to be
\begin{align}
  \vecu{P}_{i,j,k} &= \begin{pmatrix} \rho \\ \vecu{u} \\ p \\ z_n \end{pmatrix}.
\end{align}

The reconstructed state is then calculated using MUSCL reconstruction.
In this work we use the Kim-Kim 5\textsuperscript{th} order limiter
\cite{KimKim2005}, which we find to be the limiter with the least amount of
dissipation in practice, especially in the low-Mach regime. This is
consistent with the findings of Thornber \etal \cite{thornber2008lowmach}.

MUSCL reconstruction produces states associated with either side of the face.
In the context of the algorithm presented here, we just require a single state
for the face and chose this based on the sign of the
corresponding face velocity, $\uf$
\begin{align}
  \vecu{\bar{P}}_{i+1/2} = \left\{ \begin{array}{ll}
  \vecu{P}^L_{i+1/2} & \mathrm{if} \quad u_{i+1/2} > 0  \\
  \vecu{P}^R_{i+1/2} & \mathrm{otherwise}
  \end{array} \right.
\end{align}

Finally the primitive variables are converted back into conservative form,
$\vecu{\bar{U}}$ using the equation of state.

\subsection{Boundary Conditions}
\label{sec:boundary_conditions}
In this section details are given of the special handling of cell faces
corresponding to transmissive and reflective boundary conditions.

Transmissive boundary conditions are easy to handled in much the same way as
standard Riemann solver based approaches to solving the Euler equations such as
those presented by Toro \cite{Toro2009}. Ghost cells beyond the boundary are
populated with data from the interior. The ghost cell data are used directly
with the usual algorithm to calculate fluxes on the boundary of the domain. The
linear solver is set up to apply homogeneous Neumann boundary conditions.

To apply reflective boundaries we first set the predictor face velocity, $\ut$, on
the corresponding faces to zero. In event that the component of the gravity force normal to
the face of the boundary is non-zero, a corresponding inhomogeneous Neumann boundary
condition
\begin{align}
  \nabla p = \rho \vecu{g}
\end{align}
is applied, where the value of the density is taken from the interior cell
adjacent to the boundary.

Once the linear system has been solved, the ghost cells of the cell-centered
pressure field are populated in correspondence with the boundary condition,
\begin{equation}
  p_g = p_r + \rho_r \vecu{g} \cdot \hatnf \Delta x,
\end{equation}
where $g$ is used to index the ghost cell adjacent to the interior cell, $r$,
and $\hatnf$ points out of the domain. It can be easily verified that, if done
correctly, the corrected face velocity, $\uf$, calculated using
\eqref{eq:u_correction} will be zero. Population of the ghost cells in this
manner after the linear system has been solved is also necessary for the
correct determination of the face pressures, $\pf$.

Reflective boundaries can also be applied `manually' to cells within the domain
interior for the purpose of modeling embedded rigid geometries. In this case,
we can refer back to \eqref{eq:helmholtz_div} and eliminate the pressure
gradient term on the reflective cell faces by setting the corresponding face
volume terms to zero,
\begin{equation}
  \frac{1}{\rho^a_f} \leftarrow 0.
\end{equation}
For cells adjacent to reflective faces, not all cell faces are using the same
corrected velocity expression \eqref{eq:u_correction}. This means the
gravity term no longer cancels out on taking the divergence, and must now be added
for each non-reflective face. This additional term on the right hand side of
the linear system corresponds exactly to the inhomogeneous Neumann boundary
conditions that are applied for reflective boundaries at the edge of the domain.

\section{Verification and Validation}

\subsection{Shock capturing}

\subsubsection{Sod's test}

\begin{figure}
  \centering
  \includegraphics[width=\linewidth]{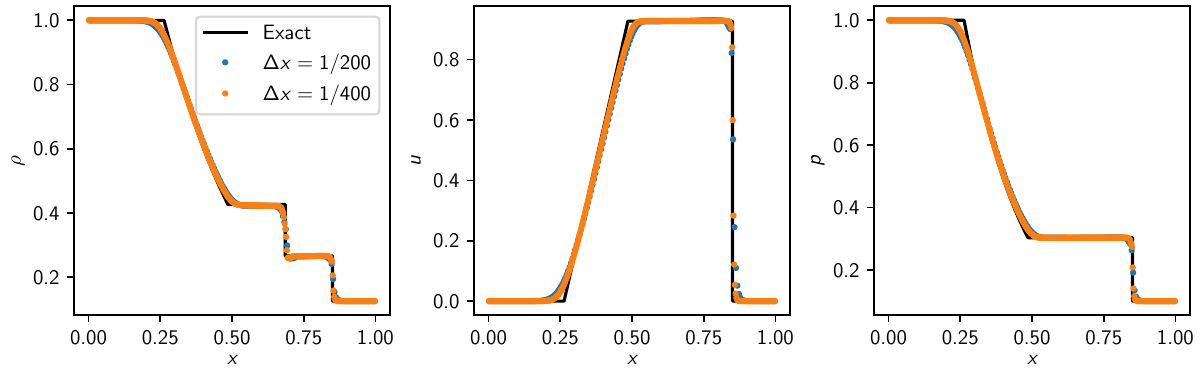}
  \caption{Results for the Sod test. The method captures shock waves
  accurately, with the correct propagation velocity and without ringing.}
  \label{fig:sod}
\end{figure}

To test the basic capability of the method to capture shock waves we first verify the
capability of the method to model the classic Sod test as described by Toro
\cite{Toro2009}. The left and right density, pressure states are
respectively $(1.0, 1.0)$ and $(0.125, 0.1)$ and the whole domain is
initialized with zero velocity. The results for plotted at a simulation time of
$0.2$, with the discontinuity initially at $0.5$ in Figure \ref{fig:sod}.
L2 errors are presented for three resolutions in Table
\ref{tab:sod_convergence}, showing that the rate of convergence is
approximately 1. This is as expected given that the slope limiters serve to
reduce the order of accuracy to $1$ in the vicinity of shock waves.

\begin{table}
{\footnotesize
  \centering
  \begin{tabular}{r|ccc}
    \hline
    $n$ & Density & Pressure & Velocity \\
    \hline
    \num{200} & \num{0.0010661 } & \num{0.0012162 } & \num{0.0028913 } \\
    \num{400} & \num{5.5637e-04} & \num{6.1364e-04} & \num{0.0013727 } \\
    \num{800} & \num{2.7549e-04} & \num{2.9385e-04} & \num{6.5862e-04} \\
    \hline
    Order of convergence
    & $0.97 \pm 0.02$
    & $1.02 \pm 0.02$
    & $1.067 \pm 0.004$ \\
  \end{tabular}
  \caption{This table shows the L2 errors for Sod's test for increasing
  resolution as calculated by comparison with the exact solution. The results
  confirm that the numerical method achieves first order convergence for shock
  waves.}
  \label{tab:sod_convergence}
}
\end{table}

\subsubsection{Circular Sod's test}
\label{sec:circular_sod}

\begin{figure}
  \centering
  \includegraphics[width=\linewidth]{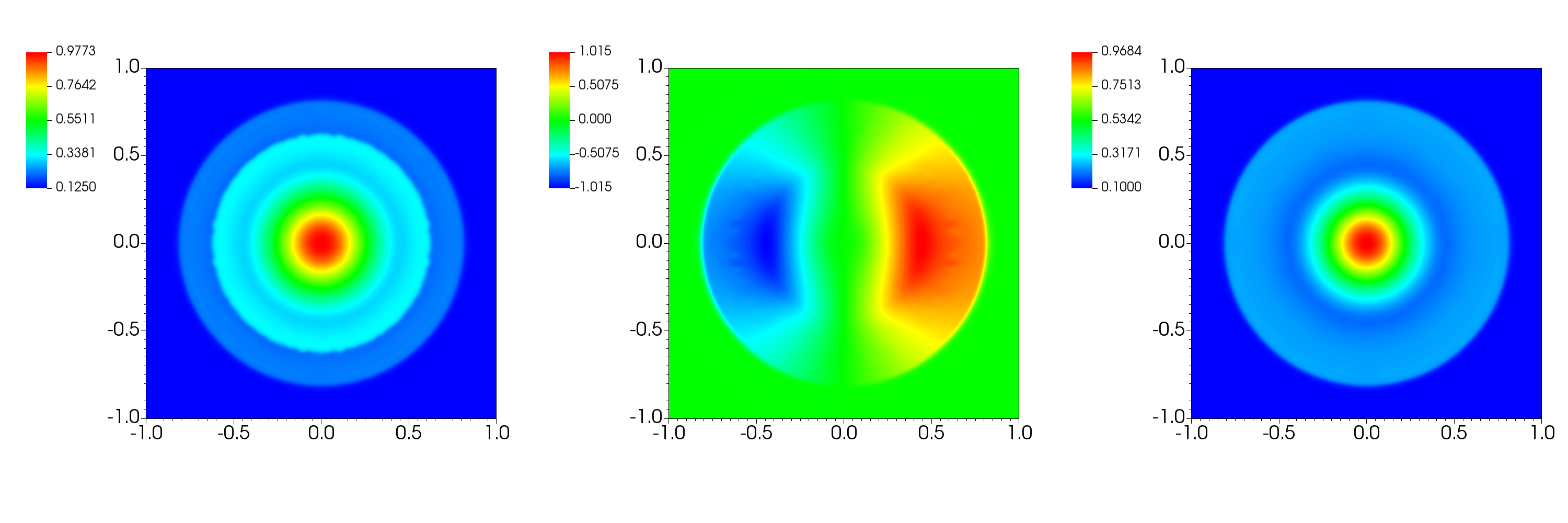}
  \caption{Two dimensional circular Sod explosion results for density, $x$
  component of the velocity vector and pressure are plotted at $0.25$ time units.}
  \label{fig:circular_sod}
\end{figure}

We also present results for the two dimensional circular Sod explosion test in
Figure \ref{fig:circular_sod}. The high pressure state $(\rho, p) = (1, 1)$ is
initialized inside a circle centered at the origin with radius $0.4$, while the
remainder of the domain is initialized with $(\rho, p) = (0.125, 0.1)$. The cell
size is $\Delta x = 0.02$.
The expression for the predictor face velocity \eqref{eq:uface_pred} uses the
average in the presence of diverging flow, and this leads to significantly improved
results in the context of diverging flow such as occurs in the circular Sod
test.

\subsubsection{Woodward-Colella blast collision test}

\begin{figure}
  \centering
  \includegraphics[width=\linewidth]{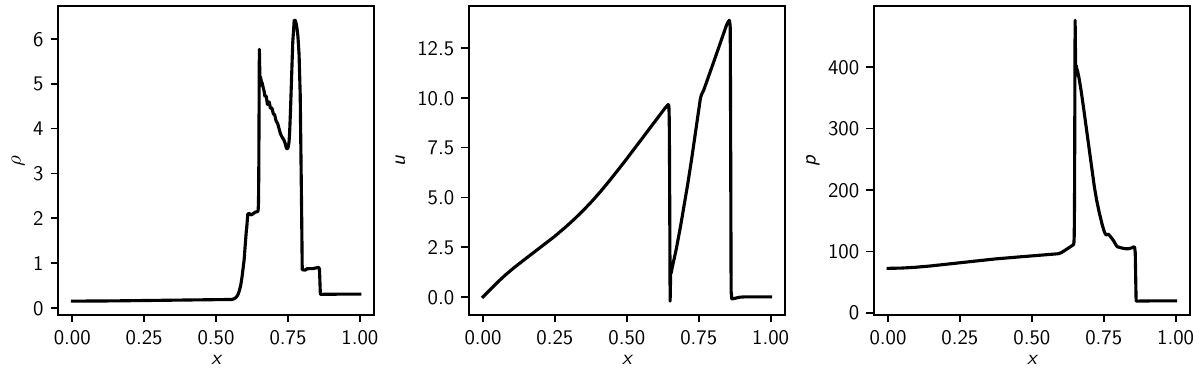}
  \caption{Results for density velocity and pressure for the blast wave
  collision test of Woodward and Colella \cite{woodward1984numerical}.}
  \label{fig:woodward_colella}
\end{figure}

Finally, the results for the classic Woodward-Colella blast wave collision test
\cite{woodward1984numerical} are shown in Figure \ref{fig:woodward_colella}. The
domain is initialized with zero velocity, density $1$ and pressure
\begin{align}
  p = \left\{ \begin{array}{ll} \num{1e3} & \text{if} \; x < 0.1 \\
    \num{1e2} & \text{if} \; 0.9 < x \\
    \num{1e-2} \quad & \text{otherwise.}
    \end{array} \right.
\end{align}
This
test highlights the robustness of the method even in the presence of very
strong shocks. The calculation uses a resolution of $\Delta x = 1/1000$, and
the results are plotted at time $0.038$.
The shocks are not as well resolved as would be expected using other schemes
designed specifically for shock capturing. Nevertheless this demonstrates the
robustness of the method presented in this work even in the presence of strong
shocks.

\subsubsection{Shock Over Backwards Step}

A Mach 1.5 shock propagates past a backwards step. After passing the step, the
shock expands radially from the corner, and a vortex develops due to the
rotational flow.

The domain is $[-5,8] \times [0,11]\;\mathrm{cm}$ with a rigid step positioned
at $[-5,1] \times [0,6]\;\mathrm{cm}$, and the ambient
state is air, $\gamma = 1.4$, density $\num{1.225e-3}\; \mathrm{kg/m^3}$,
pressure $\num{1e5} \; \mathrm{Pa}$. A Mach 1.5 shock is initialized propagating
to the right for $x < 0.5\mathrm{cm}$.

\begin{figure}
  \centering
  \includegraphics[width=\linewidth]{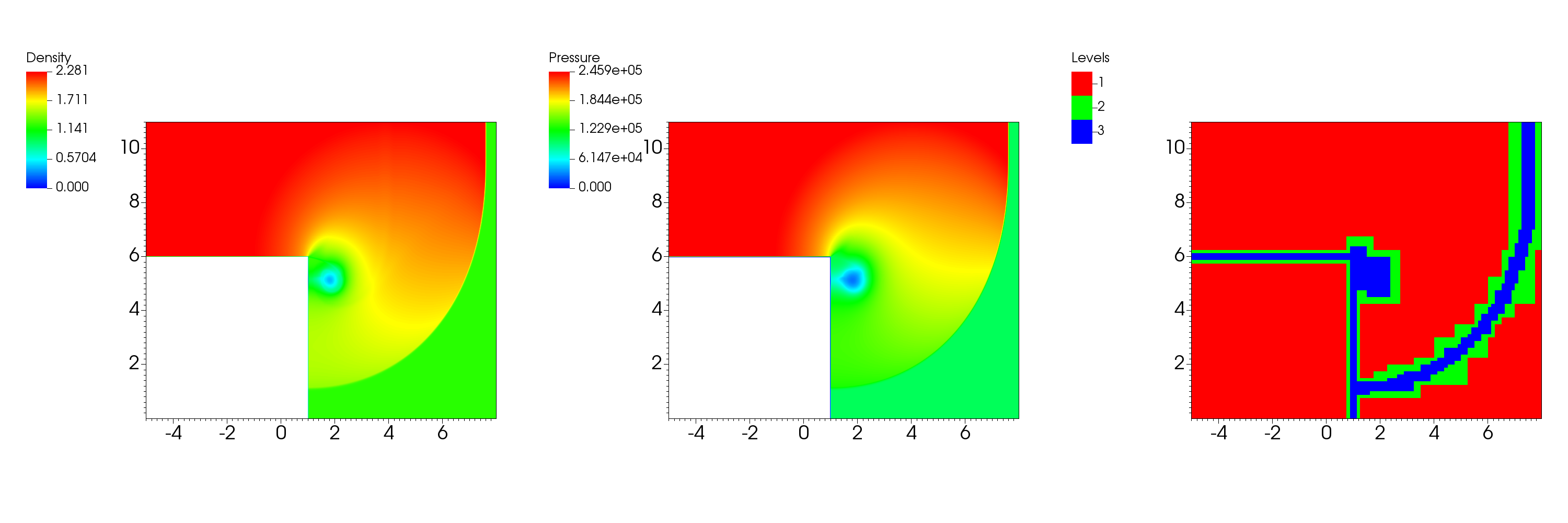}
  \caption{Results for density and pressure for the backwards step are plotted
  at $0.14\mathrm{ms}$. The shock has propagated past the step and the low
  pressure due to the vortex has developed near the edge of the step.
  The right hand plot shows the use of adaptive mesh refinement.}
  \label{fig:backwards_step}
\end{figure}

The mesh is discretized with a base cell size of $1/16$ and two levels of
factor two adaptive mesh refinement. The adaptive mesh refinement capability
comes from the AMReX library \cite{amrex} and permits
accurate capturing of shock wave waves at much lower cost than increasing the
resolution throughout the mesh. The refinement is added on the edges of the
step and over the shock wave. The areas of refinement are shown in the right
hand plot of Figure \ref{fig:backwards_step}.

We note here that in the remaining validation cases we deliberately choose not
to use AMR, since this makes it more difficult to compare the results against
other codes which do not have that capability.

\subsection{Turbulent Flow}

\subsubsection{Single mode Kelvin-Helmholtz instability}

Since we would like to model turbulent flow by way of implicit LES, we need to
establish that the numerical method is capable of resolving turbulent eddies
that are small relative to the grid size, since this determines the effective
resolution of the scheme. We model an individual Kelvin-Helmholtz vortex on a
$(16\!\times\!16)$ two dimensional mesh. This mesh is deliberately coarse and
aims to test that the method is capable of resolving the vortex even with a
very low resolution. The test is described by Thornber \etal
\cite{thornber2008lowmach}. 

For Mach number $M$, the domain of $(x,y) \in (-0.5,0.5)^2$
is initialized with
\begin{align}
  \rho &= 1 \\
  p &= 1 / (\gamma M^2) \\
  \vecu{u}(x,y) &= \left( \begin{array}{l}
      - 0.1 \sin(\pi y) \exp(-\pi \left| x \right|) \\
      \operatorname{sign}(x) (0.5 + 0.1 \cos(\pi y) \exp(-\pi \left| x \right|))
      \end{array} \right) \\
  z_0 &= \frac{1}{2} \left(1 + \operatorname{sign}(x)\right),
\end{align}
where $\vecu{u}$ has been chosen such that it is divergence free. An ideal gas
equation of state with $\gamma = 5/3$ is used. The passive scalar, $z_0$, is used
to visualize the eddy as the instability develops.

For Mach numbers $M \ll 1$, the flow is in the incompressible regime and is
expected to approach the divergence-free limit. The
resulting flow field, and tracer variable field are therefore expected to be independent of Mach number.
Meanwhile the variations in density are expected to scale like the Mach number
squared. We vary the pressure in order to impose different Mach numbers, and so
the corresponding relative variation in pressure likewise is expected to scale
like the Mach number squared.

For fully explicit numerical schemes for the Euler equations the test becomes
especially challenging as Mach number decreases, since the CFL stability
constraint reduces the timestep with decreasing Mach number, and the numerical
method can become more dissipative.
Thornber \etal \cite{thornber2008lowmach} show
that it is possible to model low-Mach flow accurately, by applying
centered differencing as opposed to upwinding to the velocity field. However the
computational cost nevertheless scales inversely with the Mach number.
In contrast, the computational cost associated with the all-Mach method
presented here is independent of Mach number.

\begin{figure}
  \centering
  \includegraphics[width=\linewidth]{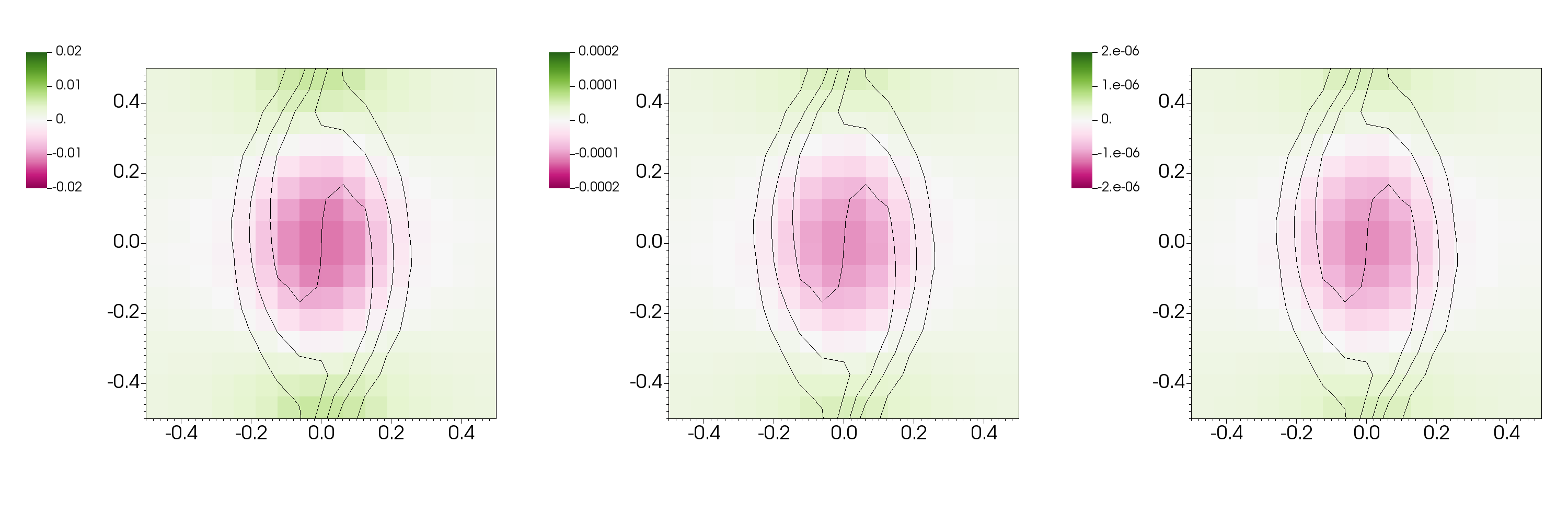}
  \caption{Density difference, $\rho - 1$, is plotted for the
  single mode Kelvin-Helmholtz test at $t = 4$. From left to right, the Mach
  numbers are $0.2$, $0.02$ and $\num{2e-3}$ respectively. The contour lines
  correspond to $z_0 = \{0.1, 0.3, 0.5, 0.7, 0.9\}$.}
  \label{fig:kh_vof}
\end{figure}

Figure \ref{fig:kh_vof} shows the density difference with respect to the
initial conditions, $(\rho - 1)$, for varying Mach number.
The overlaid contours correspond to the passive scalar, $z_0$, and show the
developing eddy of the Kelvin-Helmholtz instability. The
color map scale is set in proportion to $M^2$. We therefore expect the
resulting plot to look the same, independent of the Mach number, and this is
demonstrated in Figure \ref{fig:kh_vof}.

\begin{figure}
  \centering
  \includegraphics[width=0.6\linewidth]{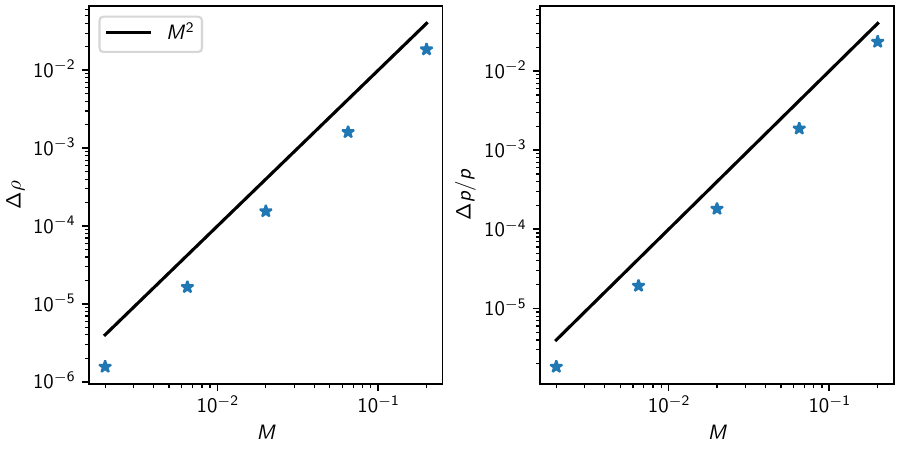}
  \caption{The variation in density and pressure scale like $M^2$ in the limit
  of small Mach number. This demonstrates that the numerical solution is
  converging towards the incompressible solution in the limit at zero Mach
  number at the physically expected rate.}
  \label{fig:kh_rho_p}
\end{figure}

The density variation, $\Delta \rho \coloneqq \max(\rho) - \min(\rho)$, and
relative variation in pressure,
$\Delta p / p \coloneqq \gamma M^2 (\max(p) - \min(p))$ are plotted as
functions of Mach number in Figure \ref{fig:kh_rho_p}. The variation scales
like $M^2$ matching the theoretical expectation. This demonstrates the
capability of the method to recover the divergence-free flow field solution in
the limit of small Mach number, even while the timestep of the numerical scheme
is independent of Mach number.

\subsubsection{Buoyant bubble in stratified atmosphere}

The rising bubble test demonstrates the capability of the model to capture
stratified material in a gravitational field and the buoyancy force acting on a
low density bubble.
This test and variations thereof have been presented by various authors. The
original version was published by Robert \cite{robert1993}. A
summary of previous publications and reference solutions for a variety of
numerical methods are presented by Andr\'es-Carcasona \etal
\cite{carcasona2023}. In this work we use the same initial conditions as
presented there, which originate from Giraldo \etal \cite{giraldo2008}.

The set up involves a perturbation to the potential temperature within an otherwise
hydrostatically stable atmosphere. All the boundaries are reflective.
The air is modeled using an ideal gas equation of state \eqref{eq:ideal_gas}.
A two dimensional rectangular domain is initialized with dimensions $x, y \in
[0, 1000] \times [0, 1500] \; \mathrm{km}$.
The atmosphere is initialized using a constant entropy (adiabatic) atmosphere,
which corresponds to a constant potential temperature, $\theta_0$. Potential
temperature, $\theta$, is defined as the temperature of the material if it were to be
isentropically expanded or compressed to match a given reference pressure,
$p_0$. It can be related to the temperature, $T = p / (\rho R)$, using
\begin{align}
  \theta_0 = T \left(\rho_0 / \rho\right)^{\gamma-1} \label{eq:potential_temperature}
\end{align}
where $R$ is the gas constant and $\rho_0 = p_0 / (R
\theta_0)$.

The pressure field is constructed by applying the hydrostatic equilibrium
condition, with acceleration due to gravity, $g$,
\begin{equation}
  \frac{\diff p}{\diff y} = -\rho g. \label{eq:hydrostatic}
\end{equation}
Analytically, this results in a pressure field,
\begin{equation}
  p(y) = p_0 \left(1 - \frac{(\gamma - 1) g y}{\gamma R \theta_0} \right)^\frac{\gamma}{\gamma - 1}.
  \label{eq:analytic_stratification}
\end{equation}
However, since the density field is discretized, a more precisely balanced
initial state can be constructed by integrating the differential equation
\eqref{eq:hydrostatic} on the mesh directly, by iteratively solving for the
subsequent cell state using
\begin{equation}
  p_i = p_{i-1} - \Delta y \frac{\rho_{i-1} + \rho_i}{2} g. \label{eq:numeric_stratification}
\end{equation}
More details of why this is more accurate are available in
\ref{sec:pressure_interpolation}.
Using \eqref{eq:potential_temperature}, we have
\begin{equation}
  p_i = \rho_i R \theta_0 (\rho_i / \rho_0)^{\gamma - 1}
\end{equation}
which can be substituted into \eqref{eq:numeric_stratification} to form a
nonlinear equation in the density, $\rho_i$.
This is solved iteratively using
Newton's method for each cell. Note that this means the initial conditions
themselves are dependent on the meshing and choice of cell size. However the
resulting field converges to the analytic solution
\eqref{eq:analytic_stratification}.

Once the hydrostatically stable pressure field, $p_i$, and corresponding
density field, $\rho_i$, has been constructed, the pressure field is maintained
while a perturbation is added to the density field. The perturbed density
field, $\tilde{\rho}_i$, is defined as
\begin{equation}
  \tilde{\rho}_i = \frac{p_i}{R \theta} 
    \left( \rho_0 / \rho \right)^{\gamma - 1}
\end{equation}
with a potential temperature field,
\begin{equation}
  \theta = \left\{ \begin{array}{lr}
    \theta_0 &  r > r_c \\
    \theta_0 + \frac{\theta_c}{2} \left(1 + \cos(\frac{\pi r}{r_c})\right) \quad & r \le r_c
  \end{array} \right.,
\end{equation}
where $r = \sqrt{(x - x_c)^2 + (y - y_c)^2}$,
$r_c = 250\;\mathrm{m}$ and $(x_c, y_c) = (500, 260) \; \mathrm{m}$.

The parameters are summarized in table \ref{tab:robert_bubble}.
\begin{table}
{\footnotesize
  \centering
  \begin{tabular}{llllll}
    \hline
    $\gamma$ & $R$ & $p_0$ & $\theta_0$ & $\theta_c$ & g \\
    \hline
    $1.4$ & $287\;\mathrm{Jkg^{-1}K^{-1}}$ & $\num{1e5}\;\mathrm{Pa}$ & $303.15\;\mathrm{K}$ & $0.5\;\mathrm{K}$ & $9.81\;\mathrm{ms^{-2}}$ \\
    \hline
  \end{tabular}
  \caption{Table of parameters for the rising bubble simulation.}
  \label{tab:robert_bubble}
}
\end{table}

The simulation can be run in 2D or 3D. In 3D, we simply update the definition
of the potential temperature field using $r = \sqrt{(x - x_c)^2 + (y - y_c)^2 +
(z - z_c)^2}$ and use $z_c = x_c$. The faster rise rate of the bubble in 3D is
compensated for by increasing the height of the domain from $1500\mathrm{m}$ to
$2500\mathrm{m}$.

\begin{figure}
  \centering
  \includegraphics[width=\linewidth]{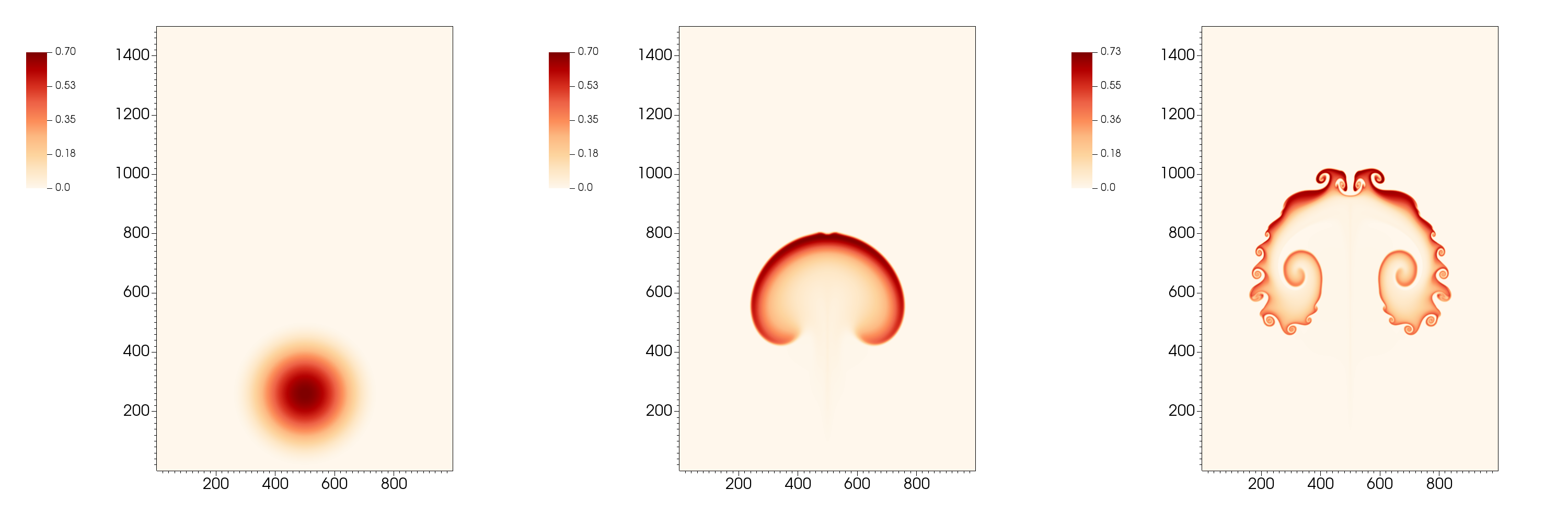}
  \caption{The evolution of the potential temperature for the thermal bubble. The initial conditions on the
  left, and plots for 9 and 13 minutes are plotted for $\Delta x = 1.25m$.}
  \label{fig:bubble_timeseries}
\end{figure}

Figure \ref{fig:bubble_timeseries} shows three time snapshots of the potential
temperature field for the 2D thermal bubble simulation with cell size, $\Delta
x = 1.25\;\mathrm{m}$.
The left hand plot
shows the initial conditions with the cosine profile perturbation on the
constant potential temperature background. The middle plot, at 9 minutes shows
the turnover of the bubble, which occurs due to the buoyant force through the
middle and the subsequent flow of material around the sides of the bubble to
fill in the space left behind. The right hand plot, at 13 minutes shows onset
of turbulence which occurs due to Rayleigh-Taylor instability at the top of the
bubble, and Kelvin-Helmholtz instability around the sides of the bubble.

\begin{figure}
  \centering
  \includegraphics[width=\linewidth]{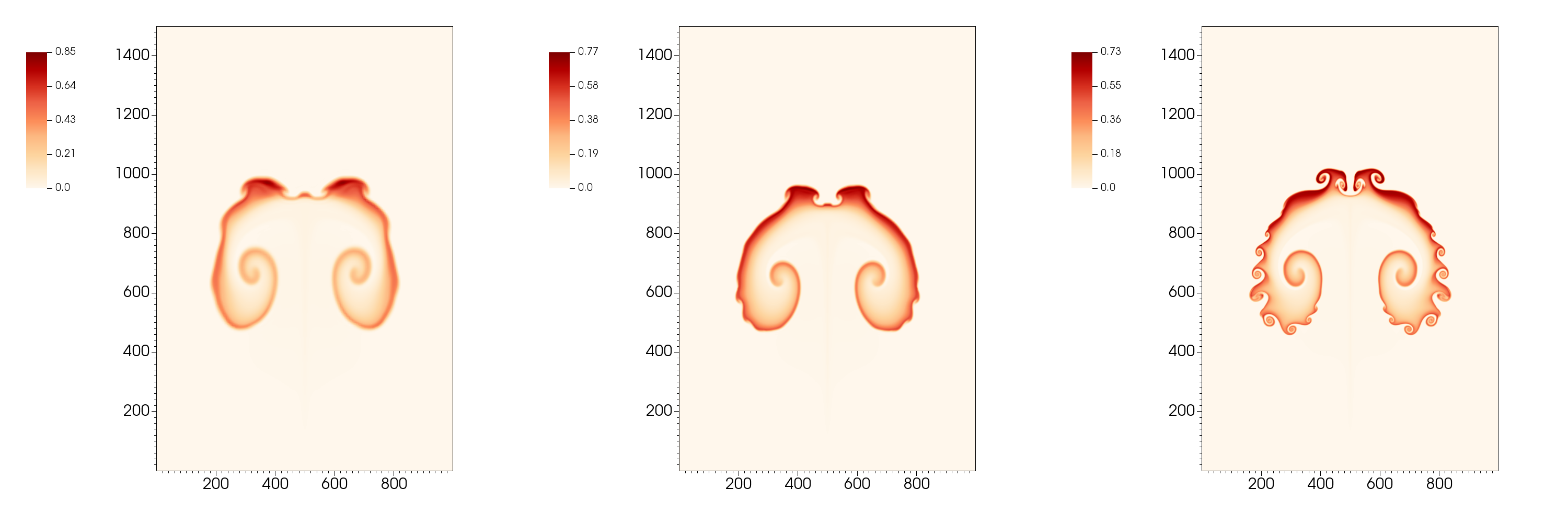}
  \caption{The onset of turbulence due to the development of Kelvin-Helmholtz
  and Rayleigh-Taylor instability is dependent on resolution. The plots show
  the potential temperature perturbation at 13 minutes using $\Delta x = \{5\mathrm{m},
  2.5\mathrm{m}, 1.25\mathrm{m}\}$ respectively.}
  \label{fig:bubble_convergence}
\end{figure}

Figure \ref{fig:bubble_convergence} shows the resolution dependence of the flow
at late time. This is expected behavior when doing ILES turbulence modeling,
since the smallest eddy which the numerical method is capable of resolving
corresponds to the resolution. Results are plotted for resolutions of
$5\;\mathrm{m}$, $2.5\;\mathrm{m}$ and $1.25\;\mathrm{m}$ illustrating that
the timing of development of Kelvin-Helmholtz instability is resolution
dependent, and thus is also in general not expected to match across different
numerical methods.

\begin{figure}
  \centering
  \includegraphics[width=\linewidth]{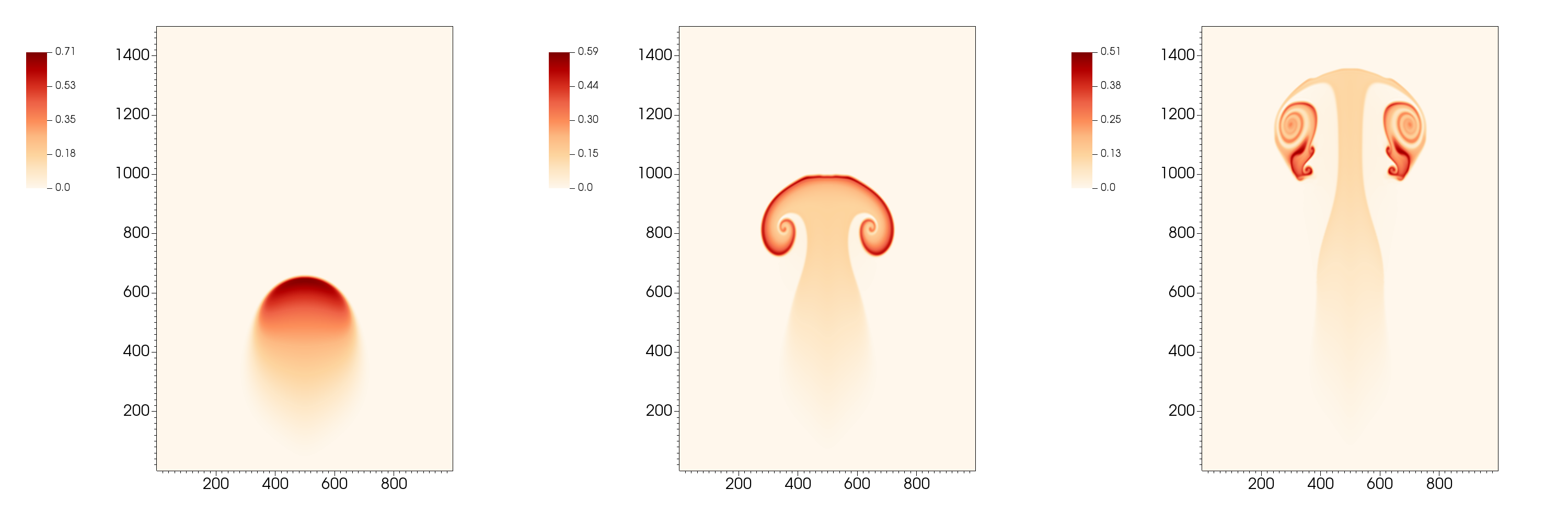}
  \caption{The potential temperature perturbation is plotted for the 3D thermal
  bubble with resolution of $\Delta x = 2.5\mathrm{m}$ is plotted from left to
  right at 5, 9 and 13 minutes respectively.}
  \label{fig:bubble_timeseries_3d}
\end{figure}

\begin{figure}
  \centering
  \includegraphics[width=0.8\linewidth]{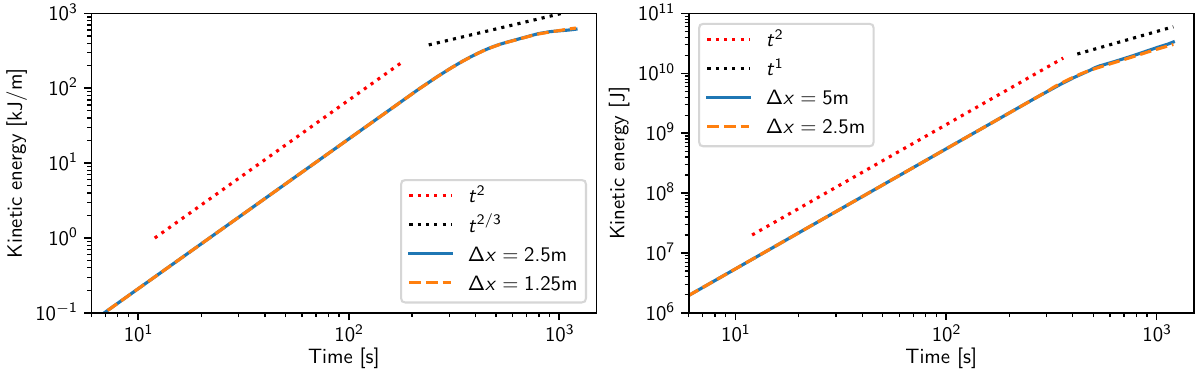}
  \caption{The development of the kinetic energy for the 2D (left) and
  3D (right) thermal bubble.
  In each case the kinetic energy initially scales with $t^2$ before an
  eventual slow down due to the onset of turbulence.}
  \label{fig:bubble_energy}
\end{figure}

Figure \ref{fig:bubble_timeseries_3d} shows time snapshots of a 2D slice
through the middle of the 3D version of the problem.
The results show favorable agreement with the results of Andr\'es-Carcasona \etal
\cite{carcasona2023}.
Figure \ref{fig:bubble_energy}
shows the evolution of the kinetic energy for the 2D (left)
and 3D (right) versions of the problem respectively.
There is an initial
period of time in which the flow is dominated by the buoyancy and the kinetic
energy in the domain scales like time squared. The flow then transitions into a
regime dominated by turbulent flow, and the kinetic energy scales like
$t^{2/3}$ in the two dimensional case and $t^1$ in the three dimensional case.
These scaling relations correspond to the results presented by Andr\'es-Carcasona \etal
\cite{carcasona2023} 

\pagebreak

\subsubsection{Richtmyer-Meshkov Simulations}

The final simulation presented in this paper is a Richtmyer-Meshkov instability
calculation. We choose the setup presented by Thornber \etal
\cite{thornber2017late}. We model turbulent mixing at the interface between a dense
and low density ideal gas. Richtmyer-Meshkov instability is challenging to model,
because the initial passage of the shock wave requires a fully compressible
code. On the other hand, after the initial rapid turbulent flow
immediately following the shock, the kinetic energy gradually decays. Later on, the
flow enters a quasi-incompressible regime, with very low Mach number, but
the width of the mixing layer continues to grow for a long time. As described
by Thornber, continuing the calculation until these late times requires huge
computational resources. This is mostly due to the CFL constraint on the
timestep for conventional compressible numerical methods, which means that at
low Mach numbers the fluid is only advected a small fraction (roughly $1/M$) of
a cell width per timestep.

The method presented in this work enables such simulations to be done at much
lower cost. The ability to model shocks, with perfect conservation, and
low-Mach number flow, and the entire range of Mach numbers occurring between the
passing of the shock and the late-time behavior, means this method is an
obvious choice.

Thornber presents a scaling law for the total kinetic energy, $E_k$, of late-time
Richtmyer-Meshkov mixing, with $\theta = 0.291$,
\begin{align}
  E_k \sim t^{3\theta -2} \approx t^{-1.127}.
\end{align}
The maximal value of magnitude of the flow velocity, will thus scale like
\begin{equation}
  u \sim t^{3\theta/2 - 1}
\end{equation}
so we expect the timestep associated with our numerical method to scale like
\begin{align}
  \Delta t(t) \sim t^{\frac{-3\theta}{2} + 1} \approx t^{0.5635}
  \label{eq:dt_progression}
\end{align}

Hence, during the late-time behavior,  the number of timesteps required to
evolve the state a time duration $t_f$ is
\begin{align}
  \#\ Timesteps \sim \int_0^{t_f} \frac{1}{\Delta t(t)} \diff t \sim t_f^{\frac{3\theta}{2}} \approx t_f^{0.4365},
\end{align}
contrasting with conventional explicit methods for which the cost will
scale linearly with $t_f$.

The domain is initialized
with an ideal gas with adiabatic gamma, $\gamma = 5/3$. The initial conditions
correspond to a Mach $1.8439$ shock wave and an Atwood number of $0.5$. They
are defined to be
\begin{equation}
  (\rho, u, p) = \left\{ \begin{array}{ll}
    (6.37497\; \mathrm{kg\,m^{-3}}, -61.48754\; \mathrm{ms^{-1}}, 399995.9\;
    \mathrm{Pa}) & \quad \mathrm{if}\;x < 3 \\
    (1 + 2 z_1(x,y,z)) \;\mathrm{kg\,m^{-3}}, -291.575\; \mathrm{ms^{-1}}, 100000 \;
    \mathrm{Pa}) & \quad \mathrm{if}\; \mathrm{otherwise}
    \end{array} \right.,
\end{equation}
where $z_1$ is a volume fraction field that defines where the interface is. It is
defined to be
\begin{equation}
  z_1(x,y,z) = \frac{1}{2} \mathrm{erfc} \left(\frac{\sqrt{\pi} (x - (3.5 + f(y,
  z)))}{ 2 \pi / 32 } \right)
\end{equation}
where $f(y,z)$ represents the perturbation added to the interface. This
construction ensures
that the interface is smoothed and does not attain step-wise features as a
result of the discretization of the mesh.

The initial perturbation applied to the interface is as initially derived by 
Youngs \cite{Youngs_iwpctm2004} 
and subsequently used by Thornber \etal \cite{thornber2017late}.
The influence of the initial perturbation on the resulting flow field is
discussed further by Youngs \cite{Youngs_iwpctm2004} and
Thornber \etal \cite{thornber2010jfm}.
The perturbation of the interface is constructed by first initializing
a two dimensional field of normally distributed random numbers, $\phi_{j,k}$,
arising from the discretization of $\phi(y,z)$ over the $y$ and $z$ directions.
Using a discrete Fourier transform, the field is filtered by applying a uniform
power spectrum, $P(k) \diff k$, to modes which correspond to wavelengths between
$\lambda_{\mathrm{min}} = L / 8$ and $\lambda_{\mathrm{max}} = L / 4$ and zero
power elsewhere. Note that since the number of modes between $k$ and $k+\diff
k$ scales like $k$, the power applied to each mode must be scaled like
$1/k$. Finally, the perturbation is scaled such that the root-mean-square
amplitude is $\lambda_{\mathrm{min}} / 10$.
The simulations presented here use the exact same perturbation field as used by
Thornber \etal \cite{thornber2017late} courtesy of Williams
\cite{williams_private_communication}.

The boundaries on the sides ($y$, $z$) of the domain are periodic.
Initially, while the compression waves are in the domain we use transmissive
boundaries.
To avoid spurious reflections of the shock and rarefaction wave off the boundaries, we start the
simulation with a domain, $[(-152/128)2\pi, 0, 0] \times [(280/128)2\pi,
2\pi, 2\pi]$. At $0.025\mathrm{s}$, once the shock and rarefaction waves have propagated a
significant distance from the interface, the domain is cut to
be slightly larger than the dimensions given by Thornber, $[0, 0, 0] \times
[(186/128)\pi, 2\pi, 2\pi]$, which cuts out the shock and rarefaction waves
entirely.  The domain is discretized using a multiple of 128 cells across the
$y$ and $z$ directions and using cubic cells.
After the domain has been cut, we replace the transmissive boundaries with 
Dirichlet boundaries, and simply populate the ghost cells with the calculated post-shock states,
\begin{equation}
  (\rho, u, p) = \left\{ \begin{array}{ll}
    (5.239400\; \mathrm{kg\,m^{-3}}, -0.07896239\; \mathrm{ms^{-1}}, 288443.8\;
    \mathrm{Pa}) & x_{\mathrm{min}} \; \text{ boundary} \\
    (1.821173\; \mathrm{kg\,m^{-3}}, -0.07896239\; \mathrm{ms^{-1}}, 288443.8\;
    \mathrm{Pa}) & x_{\mathrm{max}} \; \text{ boundary}
    \end{array} \right.. \label{eq:rm_post_shock}
\end{equation}
This ensures no spurious acoustic waves develop in the domain, and that
no net influx or outflux accumulates as the simulation progresses.

There are two ways to extend the simulation to explore further in
non-dimensional time. In Thornber \etal \cite{thornber2017late}, the length
scale of the
perturbation is reduced, thus allowing the simulation to evolve further in
non-dimensional time, while running the simulation for the same amount of real
time. The downside of this approach is that by reducing the length scale of the perturbation,
the effective resolution is also reduced.
In this work we opt to instead maintain the highest
resolution from the standard problem simulations, and simply run the simulation
for longer in real time. The domain is also lengthened to allow for the mixing layer to
continue growing without interacting with the boundaries.

The advantage of this choice is that we maintain a much higher effective
resolution, creating more confidence that the calculated flow is adequately
resolved. The downside is that the mixing layer becomes large relative to the
width and height of the domain, thereby reducing the statistical significance of
the measured width value.

The longer simulation is
initialized with a domain of $[(-284/128)2\pi,0,0] \times [(388/128)2\pi,2\pi,2\pi]$ which is then reduced at
$t=0.043\mathrm{s}$ to $[(-132/128)2\pi,0,0] \times [(288/128)2\pi,2\pi,2\pi]$.

\begin{figure}
  \centering
  \includegraphics[width=\linewidth]{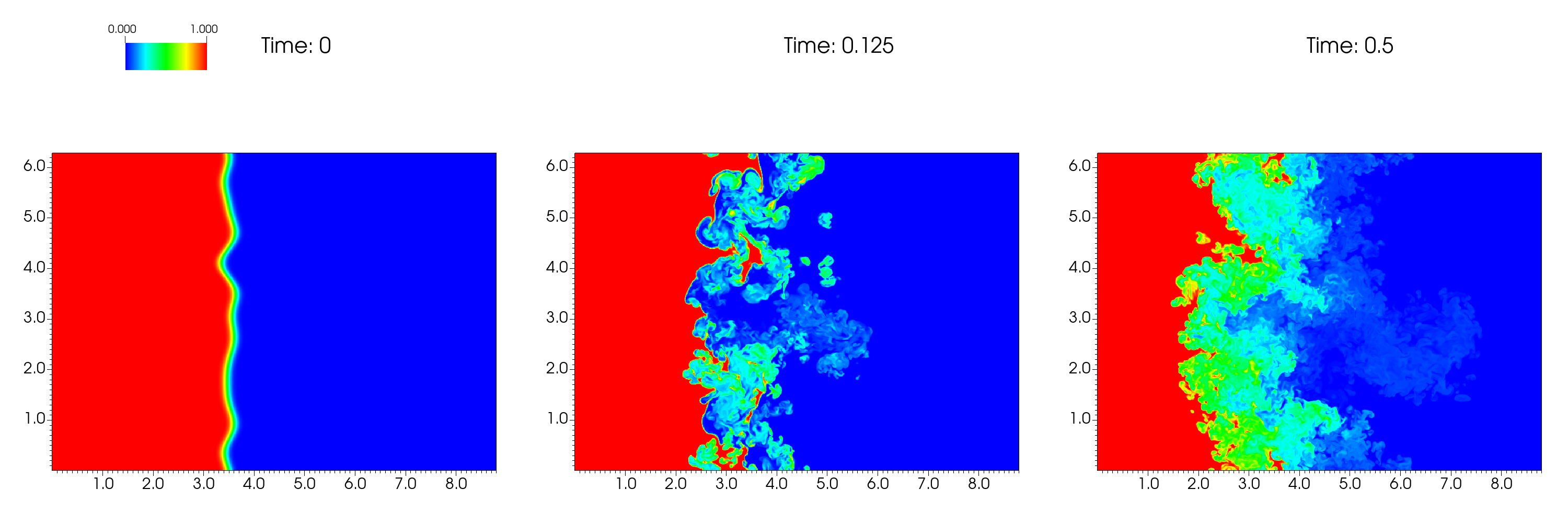}
  \vspace{0.5cm}
  \includegraphics[width=0.8\linewidth]{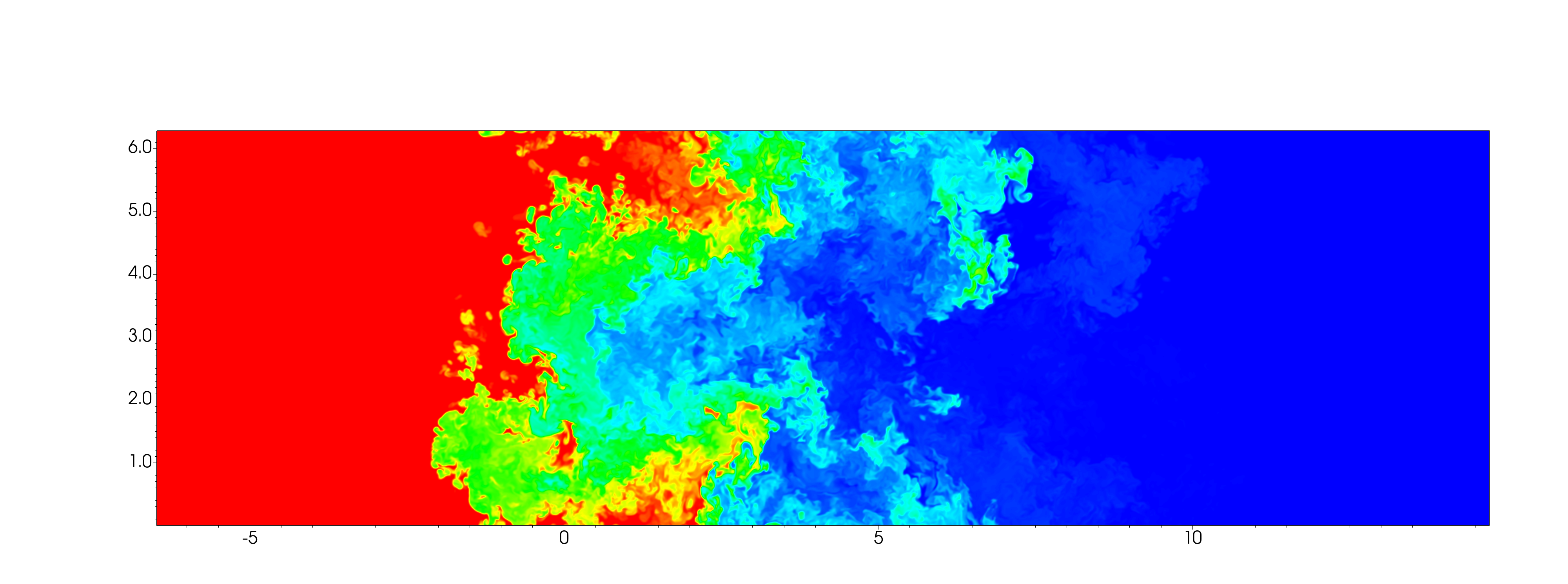}
  \caption{Slice plots of the volume fraction fields are plotted for the
  Richtmyer-Meshkov instability. The upper three plots
  show the development of the initially perturbed interface for the standard
  problem that runs to $0.5\;\mathrm{s}$. The lower plot shows the final
  volume fraction field at $20\;\mathrm{s}$ with the longer domain.}
  \label{fig:slices}
\end{figure}

Figure \ref{fig:slices} shows 2D slice plots of the volume fraction, $z_1$, for the
initial conditions and subsequent evolution for the standard problem as well as
the final volume fraction field for the late-time version of the problem.

We present results for the width of the mixing layer as a function of
time.
The mixing layer width, $W$, is defined to be
\begin{equation}
  W = \int \langle z_1 \rangle \langle (1-z_1) \rangle \diff x
\end{equation}
where $\langle \rangle$ is used to represent a planar average over $y$ and $z$.
The width is non-dimensionalized using the characteristic length scale,
$\bar{\lambda}$, which for this interface perturbation is $\frac{\pi}{4}\sqrt{\frac{12}{7}} \approx 1.028$,
while the time is non-dimensionalized using the characteristic time, $\bar{\tau} =
12.649\,\mathrm{ms^{-1}} / \bar{\lambda} \approx 12.3$. The theoretical basis for the
non-dimensionalizations is presented by Thornber \etal \cite{thornber2017late}.

\begin{figure}
  \centering
  \includegraphics[width=\linewidth]{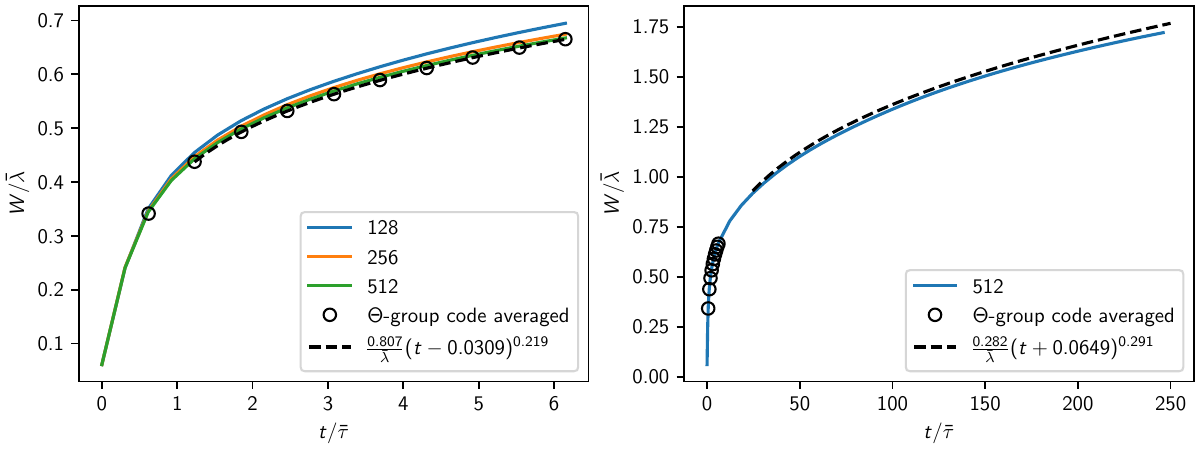}
  \caption{The mixing layer width is plotted as a function of time for the
  standard (left) and long (right) Richtmyer-Meshkov simulations. The legend
  indicates the number of cells across the domain, $N$, representing a cell
  size of $\Delta x = 2\pi / N$. The long simulation (right) is run with
  resolution corresponding to $N = 512$. The results show good agreement with the
  results from Thornber \etal \cite{thornber2017late}, although the discrepancy
  is increasing as time continues, with the mixing layer being slightly
  narrower in the simulation presented here. The code-averaged results and best fit
  curves from Thornber \etal \cite{thornber2017late} are overlaid.}
  \label{fig:mixing_layer}
\end{figure}

The left side of Figure \ref{fig:mixing_layer} shows the rate of growth of the mixing
layer as a function of time, ending at $0.5\mathrm{s}$, for comparison against
the `standard problem' results presented by Thornber \etal
\cite{thornber2017late}.
The right side of Figure \ref{fig:mixing_layer} shows the long version of the problem
continuing to a physical time of $20\;\mathrm{s}$.
The longer simulation exhibits a slightly thinner mixing layer than that
observed by Thornber, contrasting with the slightly wider mixing layer we
have measured in the early time development of the mixing layer. 

\begin{figure}
  \centering
  \includegraphics[width=0.9\linewidth]{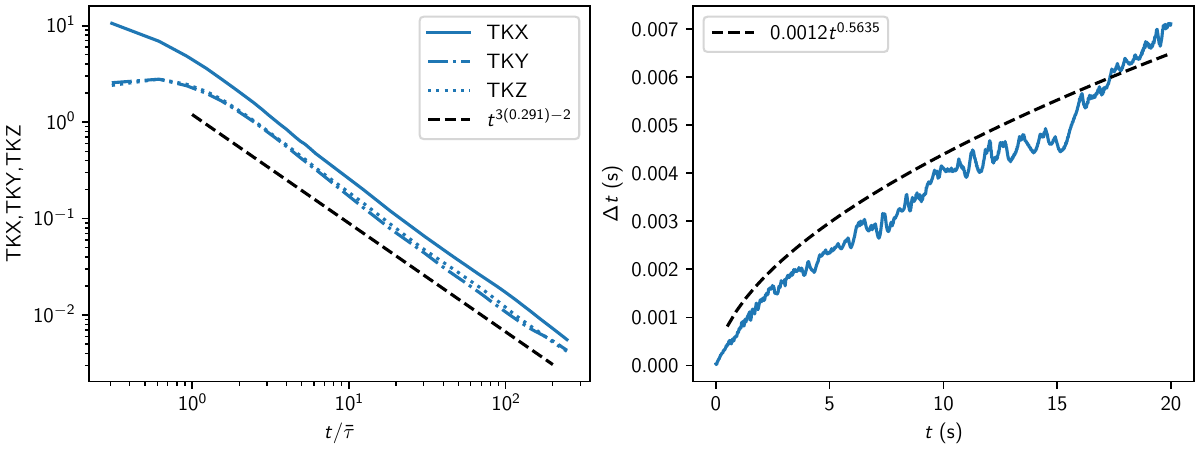}
  \caption{The left plot shows the kinetic energy scales with time in
  accordance to the expected rate throughout the longer simulation. The right
  plot shows the evolution of the computational timestep and demonstrates the
  expected increase of the timestep as the kinetic energy of the flow decays.}
  \label{fig:late_scaling}
\end{figure}

Figure \ref{fig:late_scaling} shows the decay of the kinetic energy as the simulation
progresses. 
Since the post-shock state \eqref{eq:rm_post_shock} is not perfectly stationary,
we evaluate the total kinetic energy for each direction relative to the average $x$ component of
the velocity, $\tilde{u}$,
\begin{align}
  \mathrm{TK}_i = \int \frac{1}{2} \rho (\vecu{u} \cdot \vecu{e}_i - \tilde{u})^2 \diff V,
\end{align}
for the corresponding unit vector, $\vecu{u}_i$.

The right side of Figure \ref{fig:late_scaling} shows the timestep as a function of time
for the longer Richtmyer-Meshkov simulation. The timestep is expected to scale
with the kinetic energy according to \eqref{eq:dt_progression} as a result of
the favorable stability constraints of the all-Mach method.
Consequently, the computational cost grows sublinearly with physical simulation
time. This contrasts with conventional explicit compressible schemes, for which
the timestep remains constrained by the acoustic wave speed even after the flow
has become quasi-incompressible.
As a result, the timestep continues to increase while the Mach number of the
flow decreases.
The timestep is limited by the
maximal velocity in the domain, rather than the average, which creates
fluctuations on short timescales that are not visible when examining average
kinetic energy such as is plotted on the left side of Figure
\ref{fig:late_scaling}.
This scaling of the timestep enables numerical exploration of the late-time
behavior of
Richtmyer-Meshkov instability at vastly reduced computational cost in comparison to
the methods used previously.
 
\section{Conclusions}

We have presented a novel all-Mach numerical method and validated it in
both compressible and low-Mach regimes.
Unlike hybrid compressible-incompressible approaches, the method does not
require a transition between distinct flow models, thereby avoiding sensitivity
to the choice of switching criterion.
We have validated the method for
modeling strong shocks, and verified the robustness of the algorithm
in the presence of velocity fields with large divergences. Furthermore we have
demonstrated
the ability to converge to an incompressible divergence-free flow field in
the limit of small Mach number.
We have shown that the method does not suffer from decreased accuracy in the
limit of low Mach number as can occur in compressible schemes
\cite{thornber2008lowmach},
and is thus capable of capturing turbulent eddies accurately even in the low-Mach
regime.
Finally, we have validated the applicability of the method to ILES by comparison to
Richtmyer-Meshkov Instability calculations from the literature
\cite{thornber2017late}. 

The characteristics of the method at low-Mach number are similar to many low-Mach methods
insofar as small amplitude sound waves are numerically suppressed.
As the Mach number approaches zero, the simulated flow characteristics thus
emulate the simulated flow from an incompressible method.
While all-Mach methods possess the ability to model both compression
waves with advection speed-limited timesteps, it does not constitute an
acoustic solver.

The method provides a practical route to high-resolution ILES
calculations spanning both compressible and quasi-incompressible regimes within
a single conservative formulation. This is particularly attractive for problems
such as Richtmyer-Meshkov instability, where the Mach number evolves
continuously across a wide range.

\section*{Acknowledgments}
The author acknowledges Phil Barton, Tim Law and Robin Williams for insightful
discussions, feedback and support.

\appendix

\section{Pressure Reconstruction on Cell Faces}
\label{sec:pressure_interpolation}

We calculate the face centered pressures, $p^{n+1}_{i+1/2}$, from the cell-centered
solution, $p^{n+1}$, of the linear system, \eqref{eq:helmholtz},
\begin{align}
  p^{n+1}_{i+1/2} &= \frac{\rho_{i+1} p^{n+1}_{i} + \rho_i p^{n+1}_{i+1}}{\rho_i + \rho_{i+1}}.
  \label{eq:p_reconstruction}
\end{align}

To justify this, we will show that this approach leads to the existence of a
balanced state -- meaning a state in hydrostatic equilibrium where application
of the numerical method results in zero for the face centered
velocity field, $\uf$, and a face centered pressure field satisfying
\begin{equation}
  p_{k+1/2} = p_{k-1/2} + \rho_k g \Delta z, \label{eq:discretised_hydrostatic}
\end{equation}
and hence results in a stationary state in the context of \eqref{eq:finite_volume_euler}.

Consider the following cell-centered pressure field,
\begin{align}
  p_k &= p_0 + \left(\frac{1}{2}{\rho_0} + \sum_{i=1}^{k-1} \rho_i +
  \frac{1}{2} \rho_k \right) g \Delta z , \label{eq:p_strat}
\end{align}
which is constructed on the assumption of uniform density within each cell.
This interpretation of the reconstruction within the cell is only necessary to
construct the balanced cell-centered pressure field; any
reconstruction of the density within the cell is consistent with
\eqref{eq:discretised_hydrostatic}.

Given zero initial velocity field, the predicted face velocities, $\tilde{u}_{k \pm 1/2}$, will be zero
everywhere (equation \eqref{eq:uface_pred}) meaning that the advected variables, $\rho^a$, $p^a$, will equal the
initial conditions. The averaging to cell faces, \eqref{eq:rho_face}, leads to 
\begin{align}
  \rho_{k+1/2} = \frac{\rho_k + \rho_{k+1}}{2}, \label{eq:appendix_rho_face}
\end{align}
so the evaluation of the Laplacian on the left hand side of
\eqref{eq:helmholtz} leads to 
\begin{align}
  \left(\div \frac{1}{\rho} \nabla p\right)_k &= \frac{1}{\Delta z} \left( \frac{1}{\rho_{k+1/2}}(\nabla p)_{k+1/2}
    - \frac{1}{\rho_{k-1/2}} (\nabla p)_{k-1/2} \right) \nonumber \\
    &= \frac{1}{\Delta z^2} \left( \frac{p_{k+1} - p_k}{\rho_{k+1/2}}
    - \frac{p_k - p_{k-1}}{\rho_{k-1/2}} \right),
\end{align}
which will equal zero given that from \eqref{eq:p_strat} and
\eqref{eq:appendix_rho_face}
\begin{align}
  p_{k+1} &= p_{k} + \frac{\rho_{k} + \rho_{k+1}}{2} g \Delta z \nonumber \\
  &= p_{k} + \rho_{k+1/2} g \Delta z.
\end{align}
By inspection, the linear system \eqref{eq:helmholtz} will therefore
have solution $p^{n+1} = p^a = p$ matching the initial conditions exactly.

Having recovered the pressure field exactly, it remains to be checked that the
fluxes \eqref{eq:conservative_vars} do not result in any movement in the domain.
Firstly, the face velocity correction, \eqref{eq:u_correction},
\begin{align}
  u_{k+1/2} = \tilde{u}_{k+1/2} + \Delta t \left( - \frac{p_{k+1} -
  p_k}{\rho_{k+1/2} \Delta z} + g \right),
\end{align}
will, as desired, leave the face velocities exactly zero. Secondly, we need to
to consider flux due to the pressure in the momentum equation.
Using the face pressure reconstruction scheme \eqref{eq:p_reconstruction}, we
get 
\begin{align}
  p_{k+1/2} &= \frac{\rho_{k+1} p_k + \rho_k p_{k+1}}{\rho_k + \rho_{k+1}}
  \nonumber \\
    &= \frac{\rho_{k+1} p_k + \rho_k (p_k + \rho_{k+1/2} g \Delta z)}{\rho_k + \rho_{k+1}} \nonumber \\
    &= p_k + \frac{1}{2} \rho_k g \Delta z
\end{align}
and similarly
\begin{align}
  p_{k-1/2} &= p_k - \frac{1}{2} \rho_k g \Delta z.
\end{align}
Hence, \eqref{eq:discretised_hydrostatic} is satisfied and the net update to
the momentum in the $z$ direction in \eqref{eq:finite_volume_euler} is zero.
  
Alternative approaches to reconstruction of the face pressures do not share
this property.
Practical experimentation has shown that the existence of a balanced state as
has been demonstrated here is crucial when modeling stratified atmospheres
with long timesteps to prevent the growth of unstable oscillations in the
simulation. Alternative strategies to construct the face pressures have been
found to give unsatisfactory results when used in the context of stratified flows.
Furthermore, we note that the existence of this balanced state is
dependent on the requirement that the predicted face velocities, $\ut$, are
strictly always less than or equal in magnitude to the cell-centered velocity
field. Alternative algorithms that make use of the pressure differential (such
as Riemann solvers) break this property and have proven to be problematic in
practice.

\bibliographystyle{elsarticle-num}

\bibliography{all_speed.bib}

\end{document}